\documentclass[journal]{IEEEtran}

\usepackage{amsmath,amssymb,amsfonts}
\usepackage{graphicx}
\usepackage{textcomp}
\usepackage{xcolor}
\usepackage[letterpaper, left=0.65in, right=0.65in, bottom=1in, top=0.70in]{geometry}

\def\BibTeX{{\rm B\kern-.05em{\sc i\kern-.025em b}\kern-.08em
    T\kern-.1667em\lower.7ex\hbox{E}\kern-.125emX}}
\newcommand\scalemath[2]{\scalebox{#1}{\mbox{\ensuremath{\displaystyle #2}}}}
\newcommand*{\Scale}[2][4]{\scalebox{#1}{$#2$}}%
\usepackage{xcolor}
\usepackage[linesnumbered,ruled,vlined]{algorithm2e}
\usepackage{listings}
\usepackage[noend]{algpseudocode}

\SetCommentSty{mycommfont}
\SetKwInput{KwInput}{Input}   % Set the Input
\SetKwInput{KwOutput}{Output} % set the Output
\usepackage{subcaption}
\usepackage{caption}

\definecolor{GreenForest}{rgb}{0.09, 0.45, 0.27}
\usepackage{soul}
\usepackage{balance}
\usepackage{cite}
\usepackage{acro} 
\DeclareAcronym{RIS}{
  short = RIS ,
  long  = reconfigurable intelligent surfaces ,
  class = abbrev
}
\DeclareAcronym{6G}{
  short = 6G,
  long  = 6th generation ,
  class = abbrev
}
\DeclareAcronym{5G}{
  short = 5G,
  long  = 5th generation ,
  class = abbrev
}
\DeclareAcronym{EM}{
  short = EM ,
  long  = electromagnetic,
  class = abbrev
}
\DeclareAcronym{BS}{
  short = BS,
  long  = base station ,
  class = abbrev
}
\DeclareAcronym{UE}{
  short = UE ,
  long  = user equipment ,
  class = abbrev
}
\DeclareAcronym{MISO}{
  short = MISO,
  long  = multiple-input single-output ,
  class = abbrev
}
\DeclareAcronym{MMSE}{
  short = MMSE ,
  long  = minimum mean squared error ,
  class = abbrev
}
\DeclareAcronym{RMSE}{
  short = RMSE ,
  long  = root mean squared error ,
  class = abbrev
}
\DeclareAcronym{DFT}{
  short = DFT,
  long  = discrete Fourier transform ,
  class = abbrev
}
\DeclareAcronym{FFT}{
  short = FFT,
  long  = fast Fourier transform ,
  class = abbrev
}
\DeclareAcronym{ISFFT}{
  short = ISFFT,
  long  = inverse symplectic finite Fourier transform ,
  class = abbrev
}
\DeclareAcronym{SFFT}{
  short = SFFT,
  long  = symplectic finite Fourier transform ,
  class = abbrev
}
\DeclareAcronym{THz}{
  short = THz,
  long  = Terahertz ,
  class = abbrev
}
\DeclareAcronym{IoT}{
  short = IoT,
  long  = internet of things ,
  class = abbrev
}
\DeclareAcronym{MSE}{
  short = MSE,
  long  = mean square error ,
  class = abbrev
}
\DeclareAcronym{CSI}{
  short = CSI ,
  long  = channel state information ,
  class = abbrev
}
\DeclareAcronym{MIMO}{
  short = MIMO,
  long  = multiple-input multiple-output ,
  class = abbrev
}
\DeclareAcronym{UPA}{
  short = UPA,
  long  = uniform planner array ,
  class = abbrev
}
\DeclareAcronym{RF}{
  short = RF,
  long  = radio-frequency ,
  class = abbrev
}
\DeclareAcronym{mmWave}{
  short = mmWave,
  long  = millimeter-wave ,
  class = abbrev
}
\DeclareAcronym{AoA}{
  short = AoA ,
  long  = angle of arrival ,
  class = abbrev
}
\DeclareAcronym{AoD}{
  short = AoD,
  long  = angle of departure ,
  class = abbrev
}
\DeclareAcronym{EKF}{
  short = EKF,
  long  = extended Kalman filter ,
  class = abbrev
}
\DeclareAcronym{LMS}{
  short = LMS,
  long  = least mean square ,
  class = abbrev
}
\DeclareAcronym{BiLMS}{
  short = BiLMS,
  long  = bi-directional LMS ,
  class = abbrev
}
\DeclareAcronym{SNR}{
  short = SNR,
  long  = signal-to-noise ratio ,
  class = abbrev
}
\DeclareAcronym{LoS}{
  short = LoS,
  long  = line-of-sight ,
  class = abbrev
}
\DeclareAcronym{TDD}{
  short = TDD,
  long  = time-division duplexing ,
  class = abbrev
}
\DeclareAcronym{NMSE}{
  short = NMSE,
  long  = normalized mean square error ,
  class = abbrev
}
\DeclareAcronym{PAPR}{
  short = PAPR,
  long  = peak to average power ratio ,
  class = abbrev
}
\DeclareAcronym{ISI}{
  short = ISI,
  long  = inter-symbol interference ,
  class = abbrev
}
\DeclareAcronym{ICI}{
  short = ICI,
  long  = inter-carrier interference ,
  class = abbrev
}
\DeclareAcronym{SDR}{
  short = SDR,
  long  = semidefinite relaxation ,
  class = abbrev
}
\DeclareAcronym{QoS}{
  short = QoS,
  long  = quality of service ,
  class = abbrev
}
\DeclareAcronym{NOMA}{
  short = NOMA,
  long  = non-orthogonal multiple access ,
  class = abbrev
}
\DeclareAcronym{OMA}{
  short = OMA,
  long  = orthogonal multiple access ,
  class = abbrev
}
\DeclareAcronym{NU}{
  short = NU,
  long  = near user ,
  class = abbrev
}
\DeclareAcronym{CIR}{
  short = CIR,
  long  = channel impulse response ,
  class = abbrev
}
\DeclareAcronym{FU}{
  short = FU,
  long  = far user ,
  class = abbrev
}
\DeclareAcronym{CP}{
  short = CP,
  long  = cyclic prefix ,
  class = abbrev
}
\DeclareAcronym{ZP}{
  short = ZP,
  long  = zero padding ,
  class = abbrev
}
\DeclareAcronym{ZS}{
  short = ZS,
  long  = zero suffix ,
  class = abbrev
}
\DeclareAcronym{ZF}{
  short = ZF,
  long  = zero forcing ,
  class = abbrev
}
\DeclareAcronym{RCP}{
  short = RCP,
  long  = reduced-CP ,
  class = abbrev
}
\DeclareAcronym{FZS}{
  short = FZS,
  long  = full-ZS ,
  class = abbrev
}
\DeclareAcronym{RZP}{
  short = RZP,
  long  = reduced-zero padded ,
  class = abbrev
}
\DeclareAcronym{FCP}{
  short = FCP,
  long  = full-CP ,
  class = abbrev
}
\DeclareAcronym{BER}{
  short = BER,
  long  = bit error rate ,
  class = abbrev
}
\DeclareAcronym{SIC}{
  short = SIC,
  long  = successive interference cancellation ,
  class = abbrev
}
\DeclareAcronym{PLS}{
  short = PLS,
  long  = physical layer security ,
  class = abbrev
}
\DeclareAcronym{MRT}{
  short = MRT,
  long  = maximum ratio transmission ,
  class = abbrev
}
\DeclareAcronym{AWGN}{
  short = AWGN,
  long  = additive white Gaussian noise,
  class = abbrev
}
\DeclareAcronym{SINR}{
  short = SINR,
  long  = signal-to-interference-plus-noise ratio ,
  class = abbrev
}
\DeclareAcronym{BPSK}{
  short = BPSK,
  long  = binary phase shift keying ,
  class = abbrev
}
\DeclareAcronym{QPSK}{
  short = QPSK,
  long  = quadrature phase shift keying ,
  class = abbrev
}
\DeclareAcronym{SVD}{
  short = SVD,
  long  = singular value decomposition ,
  class = abbrev
}
\DeclareAcronym{EVD}{
  short = EVD,
  long  = eigenvalue decomposition ,
  class = abbrev
}
\DeclareAcronym{PDF}{
  short = PDF,
  long  = probability density function ,
  class = abbrev
}
\DeclareAcronym{SER}{
  short = SER,
  long  = symbol error rate ,
  class = abbrev
}
\DeclareAcronym{MGF}{
  short = MGF,
  long  = moment generating function ,
  class = abbrev
}
\DeclareAcronym{2D}{
  short = 2D,
  long  = two-dimensional ,
  class = abbrev
}
\DeclareAcronym{3D}{
  short = 3D,
  long  = three-dimensional ,
  class = abbrev
}
\DeclareAcronym{CLT}{
  short = CLT,
  long  = central limit theorem ,
  class = abbrev
}
\DeclareAcronym{QAM}{
  short = QAM,
  long  = quadrature amplitude modulation ,
  class = abbrev
}
\DeclareAcronym{SISO}{
  short = SISO,
  long  = single-input single-output ,
  class = abbrev
}
\DeclareAcronym{CE}{
  short = CE,
  long  = channel estimation ,
  class = abbrev
}
\DeclareAcronym{KG}{
  short = $K_G$,
  long  = generalized-K ,
  class = abbrev
}
\DeclareAcronym{LSKRF}{
  short = LSKRF,
  long  = least squares Khatri-Rao factorization ,
  class = abbrev
}
\DeclareAcronym{FMCW}{
  short = FMCW,
  long  = frequency modulated continuous wave ,
  class = abbrev
}
\DeclareAcronym{FSK}{
  short = FSK,
  long  = frequency shift keying ,
  class = abbrev
}
\DeclareAcronym{JSAC}{
  short = JSAC,
  long  = joint sensing and communication ,
  class = abbrev
}
\DeclareAcronym{OTFS}{
  short = OTFS,
  long  = orthogonal time-frequency space ,
  class = abbrev
}
\DeclareAcronym{MP}{
  short = MP,
  long  = message passing ,
  class = abbrev
}
\DeclareAcronym{ML}{
  short = ML,
  long  = maximum likelihood ,
  class = abbrev
}
\DeclareAcronym{OFDM}{
  short = OFDM,
  long  = orthogonal frequency division multiplexing ,
  class = abbrev
}
\begin{document}
\bstctlcite{IEEEexample:BSTcontrol}
    \title{Effect of Prefix/Suffix Configurations on OTFS Systems with Rectangular Waveforms}
  \author{Salah Eddine Zegrar, and H\"{u}seyin Arslan,~\IEEEmembership{Fellow,~IEEE}

\thanks{The authors are with the Department of Electrical and Electronics Engineering, Istanbul Medipol University, Istanbul, 34810, Turkey (e-mail: salah.zegrar@std.medipol.edu.tr; huseyinarslan@medipol.edu.tr).}
% \thanks{This work has been submitted to the IEEE for possible publication. Copyright may be transferred without notice, after which this version may no longer be accessible.}
%
}
% \markboth{This work has been submitted to the IEEE for possible publication. Copyright may be transferred without notice, after which this version may no longer be accessible.}
\markboth{This work has been submitted to the IEEE for possible publication. Copyright may be transferred without notice.}%
{Shell \MakeLowercase{\textit{et al.}}: Bare Demo of IEEEtran.cls for IEEE Journals}
% ====================================================================
\maketitle

% === ABSTRACT ====================================================================
% =================================================================================
\begin{abstract}
% In this paper, we model and analyze the effective channel matrix in both time and delay-Doppler domains for orthogonal time frequency space (OTFS) systems with rectangular pulse shaping. Then we provide the input-output relation of the received signal. Different prefix/suffix configurations are considered, namely reduced-cyclic prefix (RCP), full-CP (FCP), full-zero suffix (FZS), and reduced-zero padded (RZP) OTFS. We show that the OTFS input–output relation has a simple sparse structure for all prefix/suffix types and that the only difference will be the phase term introduced when extending quasi-periodically in the delay-Doppler grid. The impact of CP length on the channel is also investigated. After that, we compare the OTFS performance using various prefix/suffix types in terms of channel estimation/equalization complexity, symbol detection performance, power and spectral efficiencies. Finally, we provide the motivation and use cases of each prefix/suffix configuration in OTFS systems and introduce a novel structure called reduced-FCP (RFCP) where the information in the CP blocks becomes extractable.

Recently, orthogonal time-frequency-space (OTFS) modulation is used as a promising candidate waveform for high mobility communication scenarios. 
In practical transmission, OTFS with rectangular pulse shaping is implemented using different prefix/suffix configurations including reduced-cyclic prefix (RCP), full-CP (FCP), full-zero suffix (FZS), and reduced-zero padded (RZP). 
However, for each prefix/suffix type, different effective channel are seen at the receiver side resulting in dissimilar performance of the various OTFS configurations given a specific communication scenario. 
To fulfill this gap, in this paper, we study and model the effective channel in OTFS systems using various prefix/suffix configurations. Then, from the input-output relation analysis of the received signal, we show that the OTFS has a simple sparse structure for all prefix/suffix types, where the only difference is the phase term introduced when extending quasi-periodically in the delay-Doppler grid.
We provide a comprehensive comparison between all OTFS types in terms of channel estimation/equalization complexity, symbol detection performance, power and spectral efficiencies, which helps in deciding the optimal prefix/suffix configuration for a specific scenario.
Finally, we propose a novel OTFS structure namely reduced-FCP (RFCP) where the information of the CP block is decodable.

\end{abstract}

% === KEYWORDS ====================================================================

\IEEEpeerreviewmaketitle

\section{Introduction}
%%%%%%%%% Intro %%%%%%%%%%%%%
\IEEEPARstart{H}{aving} a reliable wireless communication link is always a challenge due the doubly-dispersive channel effect which introduces both \ac{ISI} and \ac{ICI} \cite{matz2011fundamentals}. For instance, \ac{OFDM} overcomes the high \ac{ISI} causing frequency selectivity by dividing the whole transmission channel into smaller sub-channels in which fading is relatively considered flat \cite{ozdemir2007channel}. This is achieved by the help of \ac{CP} addition to the OFDM symbol which is longer than the delay spread of the channel. However, in time varying channels, \ac{OFDM} performs poorly due the loss of orthogonality due to the Doppler shift, leading to \ac{ICI} between the subcarriers \cite{yucek2005doppler,yusuf2016controlled}.

\Ac{OTFS} modulation has been recently introduced to parameterize the effect of the time-varying channels for any waveform by representing the channel in delay-Doppler domain where the real objects/reflectors in the propagation environment are seen, and thus, for a relatively short time frame the channel can be considered invariant \cite{hadani2017orthogonal}. The work in \cite{hadani2017orthogonal} was followed by various studies each addressing a distinct aspect of \ac{OTFS}.
For instance, \cite{mohammed2021derivation} derived and investigated the high locality of delay-Doppler signals given a limited time and bandwidth using the Zak-transform. The effect of rectangular pulse shaping on the effective channel was studied in \cite{raviteja2018practical}, where it was shown that using \ac{RCP} with OFDM modulation ensures that the delay-Doppler channel is sparse. Channel estimation and equalization were discussed in \cite{surabhi2019low}, where a low complexity \ac{MMSE} receiver is developed exploiting the sparse nature of the effective channel matrix in OTFS systems. Pilot design and detection were examined in \cite{raviteja2019embedded}, where a single pilot is embedded within the data frame protected by a two-dimensional (2D) guard in delay-Doppler domain. Then, at the receiver side after performing channel estimation, the data is detected using \ac{MP} algorithm. \ac{OTFS}-based \ac{MIMO} systems are analyzed in \cite{ramachandran2018mimo}, where the vectorized input-output relation is derived and accordingly a low complexity \ac{MP} based iterative algorithm is presented for detection. The work in \cite{tusha2022exploiting} investigated the performance of OTFS while exploiting the multi-user diversity in the downlink transmission. Finally, OTFS systems have been also studied under different hardware impairments such \ac{PAPR} \cite{francis2021diversity} and in-phase and quadrature imbalances \cite{tusha2021performance}.

%%%%%%%%%% Problem %%%%%%%%%%%%%
However, the different ways of realizing OTFS systems are neither motivated nor differentiated one from another.
For example, many research studies are based on the assumption of ideal waveform \cite{surabhi2019low,singh2021low} which is practically impossible due to Heisenberg’s uncertainty principle.
Furthermore, some studies \cite{hashimoto2021channel} neglect the phase term raised from the extension in delay-Doppler grid and assume that the data symbols in the delay-Doppler domain experience a constant channel gain, thus the effective channel matrix in delay-Doppler domain shows a doubly-circulant structure \cite{khammammetti2018otfs}. %However, practically the received signal's model is the twisted convolution between the input and the channel as explained in \cite{hadani2017orthogonal}. 
Also, when considering practical pulse shaping such as rectangular pulse shaping, many research works adopted different prefix/suffix structure. For instance, the authors in \cite{francis2021diversity,raviteja2019otfs} exploited the \ac{RCP} configuration of OTFS, while the authors in \cite{shen2019channel} used \ac{FCP} OTFS instead. In \cite{thaj2020low}, the authors chose to use a \ac{ZS} in order to combat the channel effects. However, except from the fact that the CP/ZS is appended to combat the channel dispersiveness, none of the studies mentioned above \cite{mohammed2021derivation,raviteja2018practical,surabhi2019low,raviteja2019embedded,ramachandran2018mimo,tusha2022exploiting,francis2021diversity,tusha2021performance,hashimoto2021channel,singh2021low,liu2021multi,khammammetti2018otfs,raviteja2019otfs,shen2019channel,thaj2020low} have motivated the use of a certain type of prefix/suffix or their impact on the final result of the effective channel in delay-Doppler domain. Specifically, the prefix/suffix type determines whether the channel matrix is going to be block-diagonal, block-circulant, or neither. The structure of the channel matrix in turn dictates which signal processing techniques need to be utilized.

%%%%%%%%%% Contribution %%%%%%%%%%%%%
Taking into consideration the aforementioned problem,
in this paper, the OTFS waveform is analyzed under different prefix/suffix configurations including \ac{RCP}, \ac{RZP}, \ac{FCP}, and \ac{FZS}, using rectangular pulse shaping. The conclusion of this analysis guides the researchers for the optimal OTFS design for a given scenario and system conditions. Thus, it would be easier to identify the most suitable OTFS structure based on the motivation of the study. The main contributions of this work are summarized as follows:
\begin{itemize}
    \item We first derive the input-output relationship for the \ac{RCP}, \ac{RZP}, \ac{FCP}, and \ac{FZS} OTFS systems. It is shown that the received signal has a general structure regardless of the prefix/suffix configuration used in OTFS signal structure. The received signal's structure varies in terms of the phase shift introduced when extending in the delay and Doppler grids. Also, it is deduced that in case of very low mobility or static channels, this phase term vanishes if \ac{FCP}/FZS is used, where the convolution operation converts from 2D twisted convolution to a 2D circular one.
    
    \item We provide a comprehensive performance comparison between all OTFS types in terms of channel estimation/equalization complexity, symbol detection performance, power and spectral efficiencies. Given that the same detector is used, We show that the complexity and the detection performance are the same for all configurations.
    
    % \item We show that in \ac{FCP} OTFS systems, the effective channel matrix is a function of the CP length which can be used for other purposes such as minimizing the fractional Doppler effects, rather than only being used for combating \ac{ISI}.
    
    \item Then, the comparison between the full and the reduced prefix structure is made, and it is shown that the use of \ac{RCP} OTFS is more advantageous, being more power and spectral efficient. Therefore, the use of \ac{FCP} over \ac{RCP} OTFS should be carefully selected due to the extra time and power resources needed in \ac{FCP} implementation. For instance, FCP shows superiority over other types in a system with fractional Doppler channels.
    
    \item Finally, we propose a novel CP structure for OTFS signal namely reduced-\ac{FCP} (RFCP) where RCP is appended to the FCP OTFS frame. Unlike the CP in OFDM signal, in the proposed RFCP OTFS, the CP block can be decodable instead of being discarded at the receiver side.
\end{itemize}

The rest of this paper is organized as follows; Section \ref{section:Signal-model} presents the system model used for OTFS transmission. The effect of different prefix/suffix configurations is discussed in Section \ref{Sec:CP_effect}. In Section \ref{sec:discussion}, the results are analysed and discussed. Finally, Section \ref{section:Conclusion} concludes the paper.
\footnote{\textit{Notation:} Bold uppercase $\mathbf{A}$, bold lowercase $\mathbf{a}$, and unbold letters $A,a$ denote matrices, column vectors, and scalar values, respectively. $[\alpha]_{\beta}$ takes modulo-$\beta$ of $\alpha$. $(\cdot)^H$, $(\cdot)^T$, and $(\cdot)^{-1}$ denote the Hermitian, transpose, and inverse operators. $\delta(\cdot)$ denotes the Dirac-delta function. $\operatorname{E}(\cdot)$ denotes the expectation operator. $\operatorname{diag}\left( \mathbf{A}_1,\dots,\mathbf{A}_{N} \right)$ and $\operatorname{Circ}\left(\mathbf{A}_1,\dots,\mathbf{A}_{N} \right)$ returns the block-diagonal and the block-circulant matrices composed of $\mathbf{A}_1,\dots,\mathbf{A}_{N}$, respectively. $\mathbb{C}^{{M\times N}}$ denotes the space of $M\times N$ complex-valued matrices, and $\operatorname{TriL}(\mathbf{A})$ and $\operatorname{vec}(\mathbf{A})$ are the lower triangular and the vectorized matrices of $\mathbf{A}$, respectively. $\mathbf{A} \otimes \mathbf{B}$ is the kronecker product of $\mathbf{A}$ and $\mathbf{B}$ and symbol $j$ represents the imaginary unit of complex numbers with $j^2=-1$.}

\section{System Model}\label{section:Signal-model}

\subsection{Transmitter}

Consider a wireless communication system where $M\times N$ data symbols are modulated using OTFS transform. The OTFS modulator distributes these symbols $\{X_{\mathrm{DD}}(l, k),l = 0,\dots,M-1~, ~ k=0,\dots,N-1\}$ in the 2D delay-Doppler grid which is discretized into $M$ delays and $N$ Dopplers. Then, the samples $X_{\mathrm{DD}}(l, k)$ are converted to the time–frequency domain grid $\{X_{\mathrm{TF}}(n,m),~n=0,\dots,N-1, ~ m = 0,\dots,M-1\}$ using the \ac{ISFFT} as follows 
\begin{equation}
    X_{\mathrm{TF}}(n, m)=\frac{1}{\sqrt{M N}} \sum_{l=0}^{M-1} \sum_{k=0}^{N-1} X_{\mathrm{DD}}(l, k) e^{j 2 \pi\left(\frac{n k}{N}-\frac{m l}{M}\right)}.
    \label{equ:ISFFT}
\end{equation}

To cast \eqref{equ:ISFFT} from serial to matrix-vector form, we define $\mathbf{F}_N$ to be the $N\times N$ unitary \ac{DFT} matrix; therefore, \eqref{equ:ISFFT} is rewritten as
\begin{equation}
    \mathbf{X}_{\mathrm{TF}}=\mathbf{F}_M \mathbf{X}_{\mathrm{DD}}\mathbf{F}_N^H .
    \label{equ:Xtf}
\end{equation}
Next, $\mathbf{X}_{\mathrm{TF}}$ is converted to time signal $\mathbf{s}$ by applying the Heisenberg transform. Then, $\mathbf{s}$ is transmitted over the doubly-dispersive channel as follows
\begin{equation}
\mathbf{s}=\operatorname{vec}\left(\mathbf{S}\right)= (\mathbf{F}_N^{H} \otimes \mathbf{I}_M)\mathbf{x} ,
\label{equ:time}
\end{equation}
where $\mathbf{S}=\mathbf{X}_{\mathrm{DD}}\mathbf{F}_N^H$, $\mathbf{x} =\operatorname{vec}\left(\mathbf{X}_{\mathrm{DD}}\right)$, and $\mathbf{I}_M$ denotes the $M\times M$ identity matrix.

% \begin{figure*}
%     \centering
%     \includegraphics[scale=0.75]{ CP_configuration.pdf}
%     \caption{The effect of CP configuration on the general OTFS frame: (a) Zero-CP OTFS (b) \ac{RCP} OTFS (c) \ac{FCP} OTFS.}
%     \label{fig:CP_config}
% \end{figure*}

\subsection{Channel}
Consider the $L$-tap doubly-selective \ac{CIR}, modeling $L$ propagation paths in the environment as follows   
\begin{equation}
    h(\tau,\nu) = \sum_{i=0}^{L-1}h_i\delta(\tau-\tau_i)\delta(\nu-\nu_i),
    \label{equ:channel}
\end{equation}
where $h_i$, $\tau_i = \frac{l_i}{M\Delta f}$ and $\nu_i= \frac{k_i}{NT} $ denote the complex channel gain, delay, and Doppler shift corresponding to the $i$-th path, respectively.

%  The received continuous-signal $r(t)$ can be readily expressed as \cite{raviteja2018practical}
%  \begin{equation}
% r(t)=\iint h(\tau, \nu) s(t-\tau) e^{j 2 \pi \nu(t-\tau)} d \tau d \nu+w(t),
% \label{equ:R_cont}
% \end{equation}
%  where $s(t)$ and $w(t)$ denote the continuous transmitted and noise signals, respectively. Then, The vector form of \eqref{equ:R_cont} is found as
 
 \subsection{Receiver}\label{subsec:receiver}
 The received time-frequency signal $Y(n, m)$ is converted back to delay-Doppler domain via \ac{SFFT}, as follows
 \begin{equation}
y(l, k)=\sum_{n=0}^{N-1} \sum_{m=0}^{M-1} Y(n, m) e^{-j 2 \pi\left(\frac{n k}{N}-\frac{m l}{M}\right)} .
\end{equation}
The input-output relation can be derived as 
 \begin{equation}
 \begin{aligned}
 \mathbf{y}&=\mathbf{H}_{\mathrm{eff}} \mathbf{x}+\mathbf{w}\\
&=(\mathbf{F}_N \otimes \mathbf{I}_M)\mathbf{H}(\mathbf{F}_N^{H} \otimes \mathbf{I}_M)\mathbf{x}+\mathbf{w},
 \end{aligned}
 \label{equ:y}
\end{equation}
where $\mathbf{H}_{\mathrm{eff}} \in \mathbb{C}^{M N \times M N}$ and $\mathbf{H} \in \mathbb{C}^{M N \times M N}$ denote the delay-Doppler and the time equivalent channel matrices, respectively.
 Note that the channel matrices $\mathbf{H}$ and $\mathbf{H}_{\mathrm{eff}}$ cannot be written explicitly unless the type of prefix/suffix is determined. Accordingly, the following section will explain in detail the effect of different prefix/suffix types on the effective channel in time and delay-Doppler domains, and thus on the received signal.

\begin{figure}[t]
	\centering
	\includegraphics[scale=0.90]{ 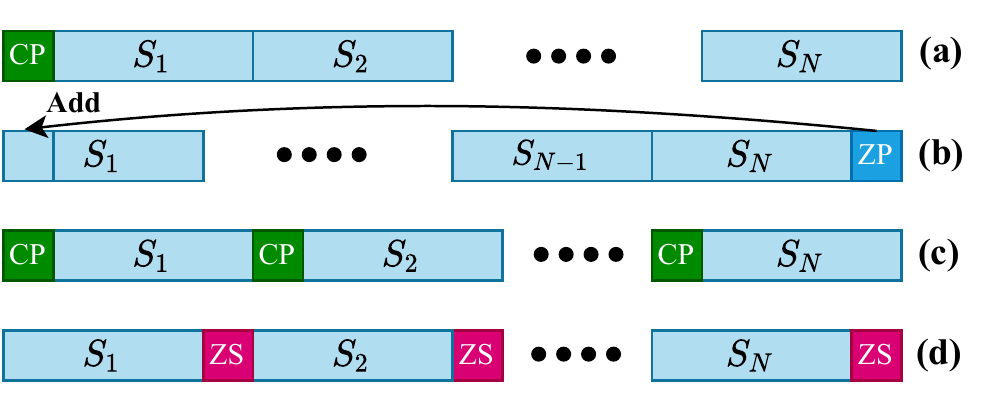}
    	\footnotesize\caption{Different prefix/suffix configurations in OTFS systems: (a) \ac{RCP} (b) \ac{RZP} (c) \ac{FCP} (d) \ac{FZS}.}
	\label{fig:prefix_conf}
\end{figure}
% \begin{figure*}
% 	\centering
% % 	\footnotesize
% 	\setcounter{mytempeqncnt}{19}
% 	\setcounter{equation}{17}
% 	\setlength{\arraycolsep}{1em}
% 	\begin{eqnarray}
% 	\begin{split}
% 	\mathbf{T}^{(i)}(p, q)= \begin{cases}e^{-j 2 \pi \frac{n}{N}} z^{k_{i}\left(\left[m-l_{i}\right]_{M}\right)}, & \text { if } q=\left[m-l_{i}\right]_{M}+M\left[n-k_{i}\right]_{N} \text { and } m<l_{i} \\ z^{k_{i}\left(\left[m-l_{i}\right]_{M}\right)}, & \text { if } q=\left[m-l_{i}\right]_{M}+M\left[n-k_{i}\right]_{N} \text { and } m \geq l_{i} \\ 0, & \text { otherwise }\end{cases}.
% 	\end{split}
% 	\end{eqnarray}
% 	\setlength{\arraycolsep}{5pt}
% 	% Restore the current equation number.
% 	\setcounter{equation}{18}
% 	% IEEE uses as a separator
% 	\hrulefill
% 	% The spacer can be tweaked to stop underfull vboxes.
% 	\normalsize
% \end{figure*}

\begin{figure*}[!t]
% ensure that we have normalsize text
\normalsize
% Store the current equation number.
%\setcounter{MYtempeqncnt}{\value{equation}}
% Set the equation number to one less than the one
% desired for the first equation here.
% The value here will have to changed if equations
% are added or removed prior to the place these
% equations are referenced in the main text.
% \setcounter{equation}{5}
\begin{equation}\tag{17}
\mathbf{T}^{(i)}(kM+l, q)= \begin{cases}e^{-j 2 \pi \frac{k}{N}} z^{k_{i}\left(\left[l-l_{i}\right]_{M}\right)}, & \text { if } q=\left[l-l_{i}\right]_{M}+M\left[k-k_{i}\right]_{N} \text { and } l<l_{i} \\ z^{k_{i}\left(\left[l-l_{i}\right]_{M}\right)}, & \text { if } q=\left[l-l_{i}\right]_{M}+M\left[k-k_{i}\right]_{N} \text { and } l \geq l_{i} \\ 0, & \text { otherwise }\end{cases}.
\label{equ:T}
\end{equation}
% Restore the current equation number.
%\setcounter{equation}{\value{MYtempeqncnt}}
% IEEE uses as a separator
\hrulefill
% The spacer can be tweaked to stop underfull vboxes.
\vspace*{4pt}
\end{figure*}
%==============================================

\section{Effect of Prefix/Suffix Configurations}\label{Sec:CP_effect}

In this section, the effect of using different types of prefix/suffix on the channel representation and received signal in OTFS systems is studied. 
Analysis starts by deriving effective channel expressions which will be used to develop the input-output relationship between the transmitted and received signals when using \ac{RCP} (Fig. \ref{fig:prefix_conf}(a)), \ac{RZP} (Fig. \ref{fig:prefix_conf}(b)), \ac{FCP} (Fig. \ref{fig:prefix_conf}(c)), and \ac{FZS} (Fig. \ref{fig:prefix_conf}(d)) configurations. Fractional Doppler shifts are not considered since the duration of one OTFS frame $NT$ is sufficient to capture the detailed channel information along the Doppler dimension over the typical wide-band systems. Hence, we assume that $l_i$ and $k_i$ are integer numbers. The case of both fractional delay and Doppler will be considered in our future work.
After each derivation, a simple example will be demonstrated where a two-tap wireless doubly-dispersive channel with \ac{CIR} $h(\tau,\nu) = h_0\delta(\tau-\frac{0}{M\Delta f})\delta(\nu-\frac{0}{NT})+h_1\delta(\tau-\frac{1}{M\Delta f})\delta(\nu-\frac{1}{NT})$ is assumed with $M= N=2$.

%==============================================
\subsection{Preliminaries}
First, we introduce the lemma on the 2D-circular convolution.

\textit{Lemma1 :} The circular convolution $C(m,n) = A(m,n)\circledast B(m,n)$ with $m= 0,\dots,M-1, ~n=0$ and $\dots,N-1$ can be expressed in a matrix form as
\begin{equation}
    \mathbf{c} = \mathbf{A}\mathbf{b},
\end{equation}
where $\mathbf{b}$ and $\mathbf{c}$ denote the vectorized form of $b(m,n)$ and $c(m,n)$, respectively. $\mathbf{A}$ is a doubly block circulant matrix generated from $a(m,n)$ with $N$ blocks, each of size $M\times M$, and it is given as
\begin{equation}
\mathbf{A}=\operatorname{Circ}\left(\mathbf{A}_0,\dots,\mathbf{A}_{N-1} \right).
\end{equation}
Each $\mathbf{A}_n$ is a $M\times M$ circulant matrix generated from the elements of the $n$-th row of $a(m,n)$, as follows
\begin{equation}
\mathbf{A}_n=\begin{pmatrix}
a(n,0) & a(n,M-1) & \dots & a(n,1) \\
a(n,1) & a(n,0) & \dots & a(n,2) \\
\vdots & \vdots & \ddots & \vdots \\
a(n,M-1) & a(n,M-2) & \dots & a(n,0)
\end{pmatrix}.
\end{equation}

\textit{Theorem 1:} The columns of the 2D \ac{ISFFT} matrix i.e., $(\mathbf{F}_N^{-1}\otimes \mathbf{F}_M)$ are eigenvectors of any doubly block circulant matrix $\mathbf{A}$. The corresponding eigenvalues are the 2D \ac{SFFT} values of the 2D signal generating the doubly block circulant matrix as follows
\begin{equation}
\boldsymbol{\Sigma}=(\mathbf{F}_N\otimes \mathbf{F}_M^{-1})\mathbf{A}(\mathbf{F}_N^{-1}\otimes \mathbf{F}_M),
\end{equation}
where $\boldsymbol{\Sigma}$ is a diagonal matrix containing the \ac{SFFT} of $a(m,n)$ generating $\mathbf{A}$.
\textit{Proof:} Please refer to Appendix \ref{App:Double_Circulant}.\hfill$\blacksquare$

\textit{Theorem 2:} $\mathbf{A}$ can be block-diagonalized as follows \cite{raviteja2018practical}
\begin{equation}
\begin{aligned}
\mathbf{A} &=\left(\mathbf{F}_{N}^{-1} \otimes \mathbf{I}_{M}\right) \mathbf{D}\left(\mathbf{F}_{N} \otimes \mathbf{I}_{M}\right) \\
&=\left(\mathbf{F}_{N} \otimes \mathbf{I}_{M}\right) \widetilde{\mathbf{D}}\left(\mathbf{F}_{N}^{-1} \otimes \mathbf{I}_{M}\right),
\end{aligned}
\end{equation}
where $\mathbf{D}$ and $\widetilde{\mathbf{D}}$ are the block diagonal matrices given in \cite{raviteja2018practical}.
%==============================================

\subsection{\ac{RCP}}
In \ac{RCP} OTFS systems, only one CP is appended to the whole \ac{OTFS} frame (see Fig. \ref{fig:prefix_conf}(a)); therefore, $\mathbf{H}^{\mathrm{R_{\mathrm{cp}}}}$ is expressed as
\begin{equation}
\mathbf{H}^{\mathrm{R_{\mathrm{cp}}}}=\sum_{i=0}^{L-1} h_{i} \boldsymbol{\Pi}^{l_{i}}_{MN} \boldsymbol{\Delta}^{k_{i}}_{MN},
\end{equation}
with $\boldsymbol{\Pi}$ being the permutation matrix,
\begin{equation}
\boldsymbol{\Pi}_{MN}=\begin{pmatrix}
0 & \cdots & 0 & 1 \\
1 & \ddots & 0 & 0 \\
\vdots & \ddots & \ddots & \vdots \\
0 & \cdots & 1 & 0
\end{pmatrix}_{M N \times M N},
\end{equation}
and $\boldsymbol{\Delta}_{MN}$ being the $M N \times M N$ diagonal matrix
\begin{equation}
\boldsymbol{\Delta}_{MN}^{k_{i}}=\operatorname{diag}\left[z_i^{0}, z_i^{1}, \ldots, z_i^{(M N-1)}\right],
\end{equation}
with $z_i=e^{\frac{j 2 \pi k_{i}}{M N}}$. 
Then, the effective channel can be found as
\begin{equation}
    \mathbf{H}_{\mathrm{eff}}^{\mathrm{R_{\mathrm{cp}}}} = (\mathbf{F}_N \otimes \mathbf{I}_M)\mathbf{H}^{\mathrm{R_{\mathrm{cp}}}}(\mathbf{F}_N^{H} \otimes \mathbf{I}_M),
\end{equation}
and it can also be expressed as \cite{raviteja2018practical}
\begin{equation}
\mathbf{H}_{\mathrm{eff}}^{\mathrm{R_{\mathrm{cp}}}}=\sum_{i=0}^{L-1} h_{i} \mathbf{T}^{(i)},
\end{equation}
where $\mathbf{T}^{(i)}$ is given in \eqref{equ:T}. Note that $\mathbf{H}_{\mathrm{eff}}^{\mathrm{R_{\mathrm{cp}}}}$ has $L-1$ non-zero elements in each row and column, reflecting the sparsity nature of the effective delay-Doppler channel matrix.
Then, the input-output relationship for each received sample $Y^{\mathrm{R_{\mathrm{cp}}}}(l, k)$ can be found as 
\begin{equation}
\setcounter{equation}{18}
\begin{aligned}
Y^{\mathrm{R_{\mathrm{cp}}}}(l, k)=\sum_{i=0}^{L-1} &h_{i}e^{ j 2 \pi\frac{ k_i}{N}\frac{ l-l_i}{M}} \Lambda_i(l,k) \\
&\times X_{\mathrm{DD}}\left(\left[l-l_{i}\right]_{M},\left[k-k_{i}\right]_{N}\right),
\end{aligned}
\label{equ:R_cont}
\end{equation}
%where $w(l, k)$ denotes \ac{AWGN}. 
where 
\begin{equation}
\Lambda_i(l, k)= \begin{cases}1 & l_{i} \leq l<M \\  e^{-j 2 \pi\frac{k}{N}} & 0 \leq l<l_{i}\end{cases}.
\end{equation}

\textit{Proof:} Please refer to Appendix \ref{App:y_Rcp}.\hfill$\blacksquare$

Since $\mathbf{H}$ is not a block-diagonal matrix $\mathbf{H}_{\mathrm{eff}}$ will not be a block-circulant \cite{raviteja2018practical}. Hence, the relationship between the data and the delay-Doppler channel response is given by the 2-D twisted convolution \cite{hadani2017orthogonal}, where a phase shift occurs as long we extend in the delay-Doppler axis. Consequently, the effective channel $\mathbf{H}_{\mathrm{eff}}$ will not be block circulant matrix. In line with our example, we find
\begin{equation}
\mathbf{H}^{\mathrm{R_{\mathrm{cp}}}}=\begin{pmatrix}
h_0 & 0 & 0 & \textcolor{black}{-jh_1} \\
h_1 & h_0 & 0 & 0 \\
0 & jh_1 & h_0 & 0 \\
0 & 0 & -h_1 & h_0
\end{pmatrix},
\label{equ:h_rcp}
\end{equation}
\begin{equation}
\mathbf{H}_{\mathrm{eff}}^{\mathrm{R_{\mathrm{cp}}}}=\begin{pmatrix}
h_0 & 0 & 0 & jh_1 \\
0 & h_0 & h_1 & 0 \\
0 & -jh_1 & h_0 & 0 \\
h_1 & 0 & 0 & h_0
\end{pmatrix}.
\label{equ:heff_rcp}
\end{equation}
As seen from \eqref{equ:h_rcp}, using one CP makes $\mathbf{H}^{\mathrm{R_{\mathrm{cp}}}}$ non-block diagonal which leads to a non-block circulant $\mathbf{H}_{\mathrm{eff}}^{\mathrm{R_{\mathrm{cp}}}}$ in \eqref{equ:heff_rcp} given that $\mathbf{H}_{\mathrm{eff}}^{\mathrm{R_{\mathrm{cp}}}}$ has non-zero elements equal to the number of channel taps in each row.

%===================================================
\subsection{\ac{RZP}}\label{Subsec:Rzp}
In \ac{RZP} OTFS systems, the whole OTFS frame is zero padded before transmission (see Fig. \ref{fig:prefix_conf}(b)). Then, at the receiver side the extra $L_{\mathrm{cp}}-1$ samples leaked from the received signal due to channel are extracted and added to the beginning of the \ac{OTFS} frame. Consequently, due to the periodic nature of $\boldsymbol{\Delta}_{MN}^{k_{i}}$ the channel is equivalent to the one considered in \ac{RCP} systems i.e., $\mathbf{H}_{\mathrm{eff}}^{\mathrm{R_{zp}}} = \mathbf{H}_{\mathrm{eff}}^{\mathrm{R_{\mathrm{cp}}}}$ thus $Y^{\mathrm{R_{zp}}}(l, k)=Y^{\mathrm{R_{cp}}}(l, k)$.

\subsection{\ac{FCP}}\label{Subsec:Fcp}
In \ac{FCP} OTFS systems, $N$ CPs of length $L_{\mathrm{cp}} \geq L$ are used to protect each OTFS subsymbol as in \ac{OFDM} modulation (see Fig. \ref{fig:prefix_conf}(c)). Different from \ac{OFDM}, where effective channel remains always the same as long as the CP duration is longer than the channel delay spread, \ac{OTFS}'s effective channel matrix is a function of \ac{CP} length in presence of Doppler. Particularly, the $(M+L_{\mathrm{cp}})N\times (M+L_{\mathrm{cp}})N$ channel matrix in this case can be given as
\begin{equation}
\hat{\mathbf{H}}^{\mathrm{F_{\mathrm{cp}}}}=\sum_{i=0}^{L-1} h_{i} \boldsymbol{\Pi}^{l_{i}}_{(M+L_{\mathrm{cp}})N} \boldsymbol{\Delta}^{k_{i}}_{(M+L_{\mathrm{cp}})N}.
\label{equ:Dopp-res}
\end{equation}

Note that when using \ac{FCP} the time duration of the OTFS frame increases, thus the Doppler resolution is higher compared to the case of \ac{RCP} i.e., $z_i = e^{\frac{j 2 \pi k_i}{(M+L_{\mathrm{cp}}) N}}\neq e^{\frac{j 2 \pi k_i}{M N}}$. This will be explained in Subsection \ref{Subsec:structure}. 

% $\tau_i = \frac{l_i}{M\Delta f}$ and $\nu_i= \frac{k_i}{NT} $; therefore,  is independent from $L_{\mathrm{cp}}$. however, since the CPs are discarded at the receiver side the  

After discarding the CPs, the channel matrix $\hat{\mathbf{H}}^{\mathrm{F_{\mathrm{cp}}}}$ becomes a block diagonal matrix with $N$ blocks, each of size $M\times M$ given as
\begin{equation}
\mathbf{H}^{\mathrm{F_{\mathrm{cp}}}}=\operatorname{diag}\left(\mathbf{H}_0^{\mathrm{F_{\mathrm{cp}}}},\dots,\mathbf{H}_{N-1}^{\mathrm{F_{\mathrm{cp}}}} \right),
\end{equation}
where
\begin{equation}
\mathbf{H}_n^{\mathrm{F_{\mathrm{cp}}}}=\sum_{i=0}^{L-1} h_{i} \boldsymbol{\Pi}^{l_{i}}_{M} \boldsymbol{\Delta}^{k_{i},n}_{M},
\label{equ:H_n_F}
\end{equation}
and 
\begin{equation}
\boldsymbol{\Delta}^{k_{i},n}_{M}=\operatorname{diag}\left[z_i^{n(M+L_{\mathrm{cp}})+L_{\mathrm{cp}}}, \dots, z_i^{((n+1)(M+L_{\mathrm{cp}})-1)}\right].
\label{equ:cp_dependency}
\end{equation}
From \eqref{equ:cp_dependency} we observe that the CP length will impact the estimated channel in the presence of Doppler. For different $L_{\mathrm{cp}}$ values, the resultant effective channel changes.
The effective channel matrix can be expressed as
\begin{equation}
    \mathbf{H}_{\mathrm{eff}}^{\mathrm{F_{\mathrm{cp}}}} = (\mathbf{F}_N \otimes \mathbf{I}_M)\mathbf{H}^{\mathrm{F_{\mathrm{cp}}}}(\mathbf{F}_N^{H} \otimes \mathbf{I}_M).
    \label{equ:H_eff_F}
\end{equation}
Applying \textit{Theorem 2}, we find
\begin{equation}
    \mathbf{H}_{\mathrm{eff}}^{\mathrm{F_{\mathrm{cp}}}} = \operatorname{Circ}\left[\mathbf{G}^{\mathrm{F_{\mathrm{cp}}}}_0, \ldots, \mathbf{G}^{\mathrm{F_{\mathrm{cp}}}}_{N-1}\right],
\end{equation}
where 
\begin{equation}
    \mathbf{G}^{\mathrm{F_{\mathrm{cp}}}}_n = \sum_{\alpha=0}^{N-1}\mathbf{H}_\alpha^{\mathrm{F_{\mathrm{cp}}}} e^{-j2\pi n\alpha/N}.
    \label{equ:G(i,j)}
\end{equation}
Each element from $\mathbf{G}^{\mathrm{F_{\mathrm{cp}}}}_n$ can be computed as the \ac{DFT} of the vector $\mathbf{g}_{\mathrm{cp}}^{(l^\prime,k^\prime)}$ conveying the $(l^\prime,k^\prime)$-th elements of $\mathbf{H}^{\mathrm{F_{\mathrm{cp}}}}_n$ i.e., $\mathbf{g}_{\mathrm{cp}}^{(l^\prime,k^\prime)} = [\mathbf{H}^{\mathrm{F_{\mathrm{cp}}}}_0(l^\prime,k^\prime),\dots,\mathbf{H}^{\mathrm{F_{\mathrm{cp}}}}_{N-1}(l^\prime,k^\prime)]^T$ as follows
\begin{equation}
\begin{aligned}
    \mathbf{G}^{\mathrm{F_{\mathrm{cp}}}}_n(l^\prime,k^\prime)& =\mathbf{f}^T_{(n,N)}\cdot\mathbf{g}_{\mathrm{zs}}^{(l^\prime,k^\prime)}= \sum_{\alpha=0}^{N-1}\mathbf{H}_\alpha^{\mathrm{F_{\mathrm{cp}}}}(l^\prime,k^\prime) e^{-j2\pi n\alpha/N}\\
    &= \sum_{\alpha=0}^{N-1}\sum_{i=0}^{L-1}h_{i}z_i^{\left(\alpha(M+L_{\mathrm{cp}})+L_{\mathrm{cp}}+l^\prime-l_i\right)}e^{-j2\pi n\alpha/N}\\
    & ~~~~~~~~~~~~~~\times \delta \left( [l^\prime-k^\prime]_M-l_{i}\right),
\end{aligned}
\label{equ:G(i,j)}
\end{equation}
where $\mathbf{f}_{(n,N)}$ denotes the $n$-th column of $\mathbf{F}_N^H$. 

\textit{Proof:} Please refer to Appendix \ref{app:H(i,j)}.\hfill$\blacksquare$

As seen from \eqref{equ:G(i,j)}, $\mathbf{G}^{\mathrm{F_{\mathrm{cp}}}}_n$ is sparse and has at most one non-zero element in each row. 
Based on \eqref{equ:y} and \eqref{equ:H_eff_F}, the input-output relationship for each received sample $Y^{\mathrm{F_{\mathrm{cp}}}}(l, k)$ for the \ac{FCP} structure can be derived as 
\begin{equation}
\begin{aligned}
Y^{\mathrm{F_{\mathrm{cp}}}}(l, k)=\sum_{i=0}^{L-1} h_{i}e^{\frac{j2\pi k_i (L_{\mathrm{cp}}+l-l_i)}{(M+L_{\mathrm{cp}})N}} X_{\mathrm{DD}}\left(\left[l-l_{i}\right]_{M},\left[k-k_{i}\right]_{N}\right).
\end{aligned}
\label{equ:y_F}
\end{equation}

\textit{Proof:} Please refer to Appendix \ref{App:y_F}.\hfill$\blacksquare$

To have further insight, we again consider our example for the case when $L_{\mathrm{cp}} = 2$. The corresponding effective channel matrices in time and delay-Doppler domains are respectively given as

\begin{equation}
\mathbf{H}^{\mathrm{F_{\mathrm{cp}}}}=\scalemath{1}{\begin{pmatrix}
h_0 & \frac{1+j}{2}h_1 & 0 & 0 \\
jh_1 & h_0 & 0 & 0 \\
0 & 0 & h_0 & -\frac{1+j}{2}h_1 \\
0 & 0 & -jh_1 & h_0
\end{pmatrix}},
\label{equ:h_fcp}
\end{equation}
and
\begin{equation}
\mathbf{H}^{\mathrm{F_{\mathrm{cp}}}}_{\mathrm{eff}}=\scalemath{1}{\begin{pmatrix}
h_0 & 0 & 0 & \frac{1+j}{\sqrt{2}}h_1 \\
0 & h_0 & jh_1 & 0 \\
0 & \frac{1+j}{\sqrt{2}}h_1 & h_0 & 0 \\
jh_1 & 0 & 0 & h_0
\end{pmatrix}}, 
\label{equ:heff_cp}
\end{equation}
As seen from \eqref{equ:h_fcp}, $\mathbf{H}^{\mathrm{F_{\mathrm{cp}}}}$ is block diagonal, thus $\mathbf{H}^{\mathrm{F_{\mathrm{cp}}}}_{\mathrm{eff}}$ is indeed block circulant with non-zero elements equal to the number of channel taps in each row as seen in \eqref{equ:heff_cp}.

\textit{Special case}: If the channel is static (i.e., $k_i = 0, ~\forall i$), the input-output relation becomes 2D circular convolution between the data and the delay-Doppler \ac{CIR} as follows
\begin{equation}
\scalemath{0.95}{Y^{\mathrm{F_{\mathrm{cp}}}}(l, k)=\sum_{i=0}^{L-1} h_{i} X_{\mathrm{DD}}\left(\left[l-l_{i}\right]_{M},\left[k-k_{i}\right]_{N}\right)=\mathbf{H}_{\mathrm{DD}}\circledast \mathbf{X}_{\mathrm{DD}}},
\label{equ:yy}
\end{equation}
where $\mathbf{H}_{\mathrm{DD}}$ represents the delay-Doppler channel matrix. According to \textit{Lemma 1} and by using \textit{Theorem 1} the received signal in \eqref{equ:yy} can be expressed as
\begin{equation}
 \begin{aligned}
 \mathbf{y}^{\mathrm{F_{\mathrm{cp}}}}&=\mathbf{C}_{\mathrm{eff}} \mathbf{x}\\
&=\left(\mathbf{F}_{N}^{-1} \otimes \mathbf{F}_{M}\right)\boldsymbol{\Sigma}\left(\mathbf{F}_{N} \otimes \mathbf{F}_{M}^{-1}\right) \mathbf{x},
 \end{aligned}
\end{equation}
where $\mathbf{C}_{\mathrm{eff}}$ is the doubly-circulant effective channel matrix. Since the channel can be simply equalized by element-wise division at the receiver side, the estimated symbols can be found as
\begin{equation}
\hat{\mathbf{x}}=\left(\mathbf{F}_{N}^{-1} \otimes \mathbf{F}_{M}\right)\boldsymbol{\Sigma}^{-1}\left(\mathbf{F}_{N} \otimes \mathbf{F}_{M}^{-1}\right)\mathbf{y}^{\mathrm{F_{\mathrm{cp}}}}.
\label{equ:lin}
\end{equation}
%==============================================
\subsection{\ac{FZS}}
In the case of \ac{FZS} OTFS, the last $L_{\mathrm{zs}}$ data symbols along the delay grid are set to zero. The \ac{FZS} can also be seen as a zero-padded signal in the delay domain (see Fig. \ref{fig:prefix_conf}(d)). The transmitted signal can be shown as
\begin{equation}
X_{\mathrm{DD}}(l,k)= \begin{cases} d_{l,k}, & \text { if } l\leq M-L_{\mathrm{zs}}-1   \\ 
0, & \text { if } l> M-L_{\mathrm{zs}}-1 \end{cases},
\end{equation}
where $d_{l,k}$ denotes the data symbols corresponding to $(l,k)$-th delay-Doppler index. The channel matrix of the \ac{FZS} $\mathbf{H}^{\mathrm{F_{zs}}}$ can be represented in terms of $\mathbf{H}^{\mathrm{F_{\mathrm{cp}}}}$ as
\begin{equation}
\mathbf{H}^{\mathrm{F_{zs}}}=\operatorname{diag}\left(\mathbf{L}_0^{\mathrm{F_{\mathrm{zs}}}},\dots,\mathbf{L}_{N-1}^{\mathrm{F_{\mathrm{zs}}}} \right),
\end{equation}
where $\mathbf{L}_\alpha^{\mathrm{F_{\mathrm{zs}}}} = \operatorname{TriL}(\mathbf{H}_\alpha^{\mathrm{F_{\mathrm{cp}}}})$.
Using the fact that $\mathbf{H}^{\mathrm{F_{zs}}}$ is block-diagonal, we apply \textit{Theorem 2} to find the effective channel matrix as follows
\begin{equation}
    \mathbf{H}_{\mathrm{eff}}^{\mathrm{F_{zs}}} = \operatorname{Circ}\left[\boldsymbol{\Omega}^{\mathrm{F_{zs}}}_0, \ldots, \boldsymbol{\Omega}^{\mathrm{F_{zs}}}_{N-1}\right],
\end{equation}
where 
\begin{equation}
    \boldsymbol{\Omega}^{\mathrm{F_{zs}}}_n = \sum_{\alpha=0}^{N-1}\mathbf{L}_\alpha^{\mathrm{F_{\mathrm{zs}}}} e^{-j2\pi n\alpha/N}.
    \label{equ:G(ii,j)}
\end{equation}

Each element from $\boldsymbol{\Omega}^{\mathrm{F_{\mathrm{zs}}}}_n$ can be computed as the \ac{DFT} of the vector $\mathbf{g}_{\mathrm{zs}}^{(l^\prime,k^\prime)}$ conveying the $(l^\prime,k^\prime)$-th elements of $\mathbf{L}^{\mathrm{F_{\mathrm{zs}}}}_n$ i.e., $\mathbf{g}_{\mathrm{zs}}^{(l^\prime,k^\prime)} = [\mathbf{L}^{\mathrm{F_{\mathrm{zs}}}}_0(l^\prime,k^\prime),\dots,\mathbf{L}^{\mathrm{F_{\mathrm{zs}}}}_{N-1}(l^\prime,k^\prime)]^T$ as follows
\begin{equation}
\begin{aligned}
    \boldsymbol{\Omega}^{\mathrm{F_{zs}}}_n(l^\prime,k^\prime) &= \sum_{\alpha=0}^{N-1}\sum_{i=0}^{L-1}h_{i}z_i^{\alpha M+k^\prime}e^{-j2\pi nk/N} \cdot \delta \left( l^\prime-k^\prime-l_{i}\right).
\end{aligned}
\label{equ:G(i,jj)}
\end{equation}

\textit{Proof:} Please refer to Appendix \ref{App:G_zs}.\hfill$\blacksquare$

The input-output relationship for the received sample $Y^{\mathrm{F_{zs}}}(l, k)$ for the \ac{FZS} structure can be calculated as 
\begin{equation}
\begin{aligned}
Y^{\mathrm{F_{zs}}}(l, k)=\sum_{i=0}^{L-1} h_{i}e^{\frac{j2\pi k_i (l-l_{i})}{MN}} X_{\mathrm{DD}}\left(l-l_{i},\left[k-k_{i}\right]_{N}\right).
\end{aligned}
\label{equ:y_Fzs}
\end{equation}

\textit{Proof:} Please refer to Appendix \ref{App:y_Fzs}.\hfill$\blacksquare$

Following up with our example, we assume $L_{\mathrm{zs}} = 1$. Then, the corresponding channel and effective channel matrices are respectively given as

\begin{equation}
\mathbf{H}^{\mathrm{F_{\mathrm{zs}}}}=\scalemath{1}{\begin{pmatrix}
h_0 & 0 & 0 & 0 \\
h_1 & 0 & 0 & 0 \\
0 & 0 & h_0 & 0 \\
0 & 0 & -h_1 & 0
\end{pmatrix}},
\label{equ:h_fzs}
\end{equation}
and
\begin{equation}
\mathbf{H}^{\mathrm{F_{\mathrm{zs}}}}_{\mathrm{eff}}=\scalemath{1}{\begin{pmatrix}
h_0 & 0 & 0 & 0 \\
0 & 0 & h_1 & 0 \\
0 & 0 & h_0 & 0 \\
h_1 & 0 & 0 & 0
\end{pmatrix}}, 
\label{equ:heff_zs}
\end{equation}
As seen from \eqref{equ:h_fzs}, $\mathbf{H}^{\mathrm{F_{\mathrm{zs}}}}$ is block diagonal with each block being lower triangular matrix, thus $\mathbf{H}^{\mathrm{F_{\mathrm{zs}}}}_{\mathrm{eff}}$ is indeed block circulant as seen in \eqref{equ:heff_zs}. Similar to the \ac{FCP} case, when the channel is static, the input-output relation is also 2D circular convolution between the data and the delay-Doppler \ac{CIR}. 
% \begin{equation}
% \begin{aligned}
% Y^{\mathrm{F_{zs}}}(l, k)=\mathbf{H}_{\mathrm{DD}}\circledast \mathbf{X}_{\mathrm{DD}}.
% \end{aligned}
% \end{equation}

%==============================================

\section{Discussions And Results}\label{sec:discussion}
In this section, we analyze and compare the different prefix/suffix OTFS waveforms based on the derivations obtained above. The comparison is made based on the received signal's model, computational complexity of channel estimation/equalization, spectral and power efficiencies.    

\subsection{The Relation Between Different OTFS Configuration}\label{Subsec:Relation}

From Section \ref{Sec:CP_effect}, it is concluded that the received signal has a common form given as
\begin{equation}
\begin{aligned}
Y(l, k)=\sum_{i=0}^{L-1} h_{i}\Gamma^{(i)}_{\eta} X_{\mathrm{DD}}\left[\left(l-l_{i}\right)_{M},\left(k-k_{i}\right)_{N}\right],
\end{aligned}
\label{equ:common}
\end{equation}
where $\Gamma^{(i)}_{\eta}(l,k,M,N,L_{\mathrm{cp/zs}})$ is a known phase term and ${\eta}$ denotes the type of prefix/suffix used $ \eta = \{\text{RCP},\text{RZP},\text{FCP},\text{FZS}\}$. Therefore, finding the relation between the $\Gamma^{(i)}_{\eta}$ terms defines the correlation between the different OTFS realizations. 

\subsubsection{\ac{RCP} \& \ac{RZP}}
It is already shown in Subsection \ref{Subsec:Rzp} that both \ac{RCP} and RZP share the same received signal's structure.

\subsubsection{\ac{RCP} \& \ac{FCP}}
Consider a \ac{RCP} OTFS system with $(M+L_{\mathrm{cp}})\times N$ data grid in delay-Doppler domain, then
\begin{equation}
   \Gamma^{(i)}_{\text{RCP}} =  e^{ j 2 \pi\frac{ k_i}{N}\frac{ l-l_i}{(M+L_{\mathrm{cp}})}} \begin{cases}1 & l_{i} \leq l<(M+L_{\mathrm{cp}}) \\  e^{-j 2 \pi\frac{k}{N}} & 0 \leq l<l_{i}\end{cases}.
\end{equation}
If the delays larger or equal to the CP length are selected (i.e., $\hat{l}=l+L_{\mathrm{cp}}\geq L_{\mathrm{cp}}$),
the condition $l_{i} \leq \hat{l}<(M+L_{\mathrm{cp}})$ is satisfied. Therefore, 
\begin{equation}
\begin{aligned}
  \Gamma^{(i)}_{\text{RCP}}&(l+L_{\mathrm{cp}},k,M+L_{\mathrm{cp}},N) =  e^{ j 2 \pi\frac{ k_i}{N}\frac{ \hat{l}-l_i}{(M+L_{\mathrm{cp}})}}\\ &= e^{ j 2 \pi\frac{ k_i}{N}\frac{L_{\mathrm{cp}}+l-l_i}{(M+L_{\mathrm{cp}})}}
  = \Gamma^{(i)}_{\text{FCP}}(l,k,M,N,L_{\mathrm{cp}}).
\end{aligned}
\end{equation}

It is concluded that using \ac{FCP} OTFS is equivalent to transmitting \ac{RCP} OTFS with parameters $\hat{M}=M+L_{\mathrm{cp}}$ and $\hat{N}=N$, and considering only $M\times N$ symbols at the receiver side corresponding to indices $l = L_{\mathrm{cp}},\dots,M+L_{\mathrm{cp}}-1$ and $k=0,\dots,N-1$.

\subsubsection{\ac{RCP} \& \ac{FZS}}

Consider a conventional \ac{RCP} OTFS system with $M\times N$ data symbols in delay-Doppler domain; however, we set the last $(M-L_{\mathrm{zs}})\times N$ symbols to zero as follows
\begin{equation}
X_{\mathrm{DD}}(l,k)= \begin{cases} d_{l,k}, & \text { if } l\leq M-L_{\mathrm{zs}}-1   \\ 
0, & \text { if } l> M-L_{\mathrm{zs}}-1 \end{cases}.
\label{equ:cond}
\end{equation}
If condition $0 \leq l<l_{i}$ is true, then
\begin{equation}
\Scale[0.95]{X_{\mathrm{DD}}\left(\left[l-l_{i}\right]_{M},\left[k-k_{i}\right]_{N}\right) = X_{\mathrm{DD}}\left(l-l_{i}+M,\left[k-k_{i}\right]_{N}\right) = 0}.
\label{equ:data}
\end{equation}
This means that for $~0 \leq l<l_{i}$, $\Gamma^{(i)}(l,k)$ can take any value since it is multiplied by zero according to \eqref{equ:data}. Then, we can generalize the phase term to be
\begin{equation}
\begin{aligned}
   \Gamma^{(i)}_{\text{RCP}}(l,k,M,N,\eqref{equ:cond}) &=  e^{ j 2 \pi\frac{ k_i}{N}\frac{ l-l_i}{M}},~~~\forall l\\ 
   &= \Gamma^{(i)}_{\text{FZS}}(l,k,M,N,L_{\mathrm{zs}}).
\end{aligned}
\end{equation}
Therefore, we conclude that using \ac{FZS} OTFS is equivalent to transmitting \ac{RCP} OTFS with parameters $M$ and $N$, and setting symbols corresponding to indices $l =M- L_{\mathrm{zs}},\dots,M-1$ and $k=0,\dots,N-1$ to zero.

\begin{figure}[t]
	\centering
	\includegraphics[scale=0.54]{ 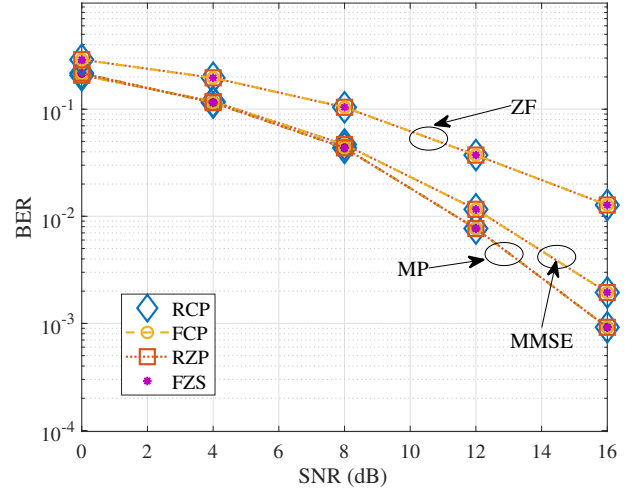}
    	\footnotesize\caption{Performance comparison of MP, ZF, and MMSE detectors based OTFS with different prefix/suffix types using $(M;N;L_{\mathrm{cp}},L_{\mathrm{zs}})=(16,16,4,4)$. The number of paths in the delay-Doppler domain is $L = 4$.}
	\label{fig:CE}
\end{figure}

% \begin{figure*}[!h]
%     \begin{center}
%     % \centering
%     \setcounter{figure}{4}
%     \subfloat[$L_{\mathrm{cp}} = 4$. ]{\label{convPerf:1}\includegraphics[width=63mm]{ F_cp_Doppler_4.eps}}
%     \subfloat[$L_{\mathrm{cp}} = 5$.]{\label{convPerf:2}\includegraphics[width=63mm]{ F_cp_Doppler_5.eps}}
%     \subfloat[$ L_{\mathrm{cp}} = 6$.]{\label{convPerf:2}\includegraphics[width=63mm]{ F_cp_Doppler_6.eps}}
%     \\
%     \end{center}
%     \centering
%     \caption{The delay-Doppler \ac{CIR} of \ac{FCP} OTFS under fractional Doppler using different CP lengths. The number of paths in the delay-Doppler domain is $L = 4$.}
%     \label{fig:CP_dep}
% \end{figure*}

\subsection{Channel Estimation}\label{Subsec:channel}

From Subsection \ref{Subsec:Relation}, we found out that regardless of the prefix/suffix type, the received signal has a common formula expressed by \eqref{equ:common}. Therefore, we can conclude that the computational complexity and the performance of a specific channel estimation/equalization algorithm and detection method are the same no matter what prefix/suffix configuration is used in OTFS systems\footnote{Note that this statement is correct if and only if the same estimator/detector is used for comparison.}. 
Three detectors are considered in the simulations, namely, the \ac{MP} detector proposed in \cite{raviteja2018interference}, \ac{ZF}  receiver \cite{singh2022ber} expressed by
\begin{equation}
\mathbf{x}_{\mathrm{ZF}}=\mathbf{H}_{\mathrm{eff}}^{H}\left(\mathbf{H}_{\mathrm{eff}} \mathbf{H}_{\mathrm{eff}}^{H}\right)^{-1},
\end{equation}
and the classic \ac{MMSE} detector given by \cite{surabhi2019low}
\begin{equation}
\mathbf{x}_{\mathrm{MMSE}}=\mathbf{H}_{\mathrm{eff}}^{H}\left(\mathbf{H}_{\mathrm{eff}} \mathbf{H}_{\mathrm{eff}}^{H}+\sigma^{2} \mathbf{I}\right)^{-1},
\end{equation}
where $\sigma^{2}$ denotes the \ac{AWGN} noise variance.
The results are depicted in Fig. \ref{fig:CE}, where it is seen that the average \ac{BER} performances of \ac{RCP}, \ac{RZP}, \ac{FCP}, and FSZ are almost identical when \ac{MP}, ZF, or \ac{MMSE} detectors are implemented for symbol detection. MMSE detector performs better than ZF since it exploits noise statistics in detection, whereas MP outperforms both because it exploits the noise statistics as well as the diversity from different channel taps.

\subsection{Spectral \& Power Efficiencies}\label{Subsec:PSE}

The total spectral and power efficiencies in OTFS systems depend directly on the prefix/suffix configuration used. To show that, consider the capacity $C$ for a given instantaneous \ac{SNR} $\gamma$ as follows \cite{liu2021optimizing}
% \begin{subequations}
%      \begin{align}\label{eq:const1}
%       & C_{R_{\mathrm{cp}}} =\left(\frac{MN}{MN+L_{\mathrm{cp}}}\right) \log _{2}\left(1+\gamma\right) \\ \label{eq:const2}
%       & C_{R_{\mathrm{zp}}} =\left(\frac{MN}{MN+L_{\mathrm{cp}}}\right) \log _{2}\left(1+\gamma\right)\\ \label{eq:const3}
%       & C_{F_{\mathrm{cp}}} = \left(\frac{M}{M+L_{\mathrm{cp}}}\right) \log _{2}\left(1+\gamma\right)\\\label{eq:const4}
%       & C_{F_{\mathrm{zs}}} = \left(\frac{M-L_{\mathrm{cp}}}{M}\right) \log _{2}\left(1+\gamma\right).
%      \end{align}
%      \label{equ:const}
% \end{subequations}
\begin{subequations}
     \begin{align}\label{eq:const1}
      & C_{R_{\mathrm{cp}}} =\left(\frac{MN}{MN+L_{\mathrm{cp}}}\right) \log _{2}\left(1+\gamma\right) \\ \label{eq:const2}
      & C_{R_{\mathrm{zp}}} =\left(\frac{MN}{MN+L_{\mathrm{cp}}}\right) \log _{2}\left(1+\gamma\right)\\ \label{eq:const3}
      & C_{F_{\mathrm{cp}}} = \left(\frac{M}{M+L_{\mathrm{cp}}}\right) \log _{2}\left(1+\gamma\right)\\\label{eq:const4}
      & C_{F_{\mathrm{zs}}} = \left(\frac{M-L_{\mathrm{zs}}}{M}\right) \log _{2}\left(1+\gamma\right).
     \end{align}
     \label{equ:const}
\end{subequations}

\begin{figure}[t]
	\centering
	\includegraphics[scale=0.50]{ 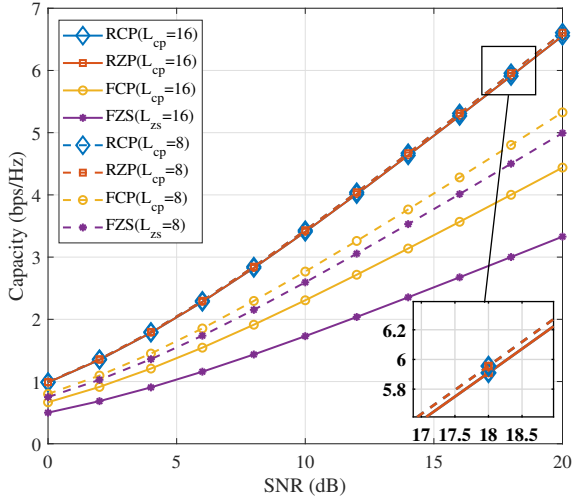}
    	\footnotesize\caption{OTFS system's capacity using different prefix/suffix configurations with $(M;N)=(32,32)$.}
	\label{fig:Capacity}
\end{figure}

\begin{figure}[t]
	\centering
	\includegraphics[scale=0.50]{ 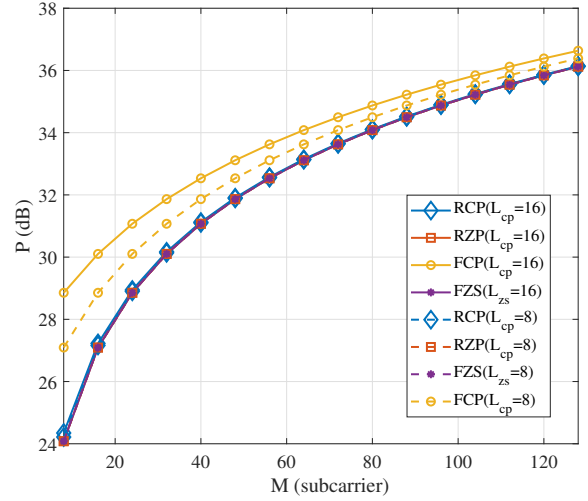}
    	\footnotesize\caption{The transmitted power required in OTFS system versus the total number of subcarriers $M$ for a given capacity and a fixed $N=32$.}
	\label{fig:PE}
\end{figure}

Regarding the power efficiency, the average transmitted power is considered as follows \cite{wulich2005definition} 
\begin{equation}
P = E\left\{\int_{0}^{T^{\prime}}|s(t)|^{2} d t\right\} =\int_{0}^{T^{\prime}}E\left\{|s(t)|^{2} \right\} d t,
\end{equation}
where $T^\prime$ is the transmitted frame's length and $E\left\{|s(t)|^{2} \right\}$ is the average symbol power. Therefore, under the constraint that all OTFS configurations have the same capacity, the average transmitted powers for each OTFS configuration are given as
\begin{subequations}
     \begin{align}\label{eq:Pconst1}
      & P_{R_{\mathrm{cp}}} =\left(MN+L_{\mathrm{cp}}\right) E\left\{|s(t)|^{2} \right\} \\ \label{eq:Pconst2}
      & P_{R_{\mathrm{zp}}} =\left(MN\right) E\left\{|s(t)|^{2} \right\}\\ \label{eq:Pconst3}
      & P_{F_{\mathrm{cp}}} = N\left(M+L_{\mathrm{cp}}\right) E\left\{|s(t)|^{2} \right\}\\\label{eq:Pconst4}
      & P_{F_{\mathrm{zs}}} = \left(MN\right) E\left\{|s(t)|^{2} \right\}.
     \end{align}
     \label{equ:Pconst}
\end{subequations}
As seen from \eqref{equ:const} and \eqref{equ:Pconst}, \ac{RZP} OTFS achieves the best attainable performance in terms of both metrics, where \ac{RCP} OTFS losses some power efficiency due to its CP but this would provide better synchronization performance and smoother transition between the OTFS frames. The last two types namely, \ac{FZS} and \ac{FCP} OTFS have the worst power and spectral efficiencies, respectively. 
Fig. \ref{fig:Capacity} and Fig. \ref{fig:PE} emphasizes the above results. For instance, Fig. \ref{fig:Capacity} shows that RCP and RZP provide the highest capacities which are almost invariant for the cases of $L_{\mathrm{cp}}=8$ and $L_{\mathrm{cp}}=16$. In fact, the zoomed part of Fig. \ref{fig:Capacity} illustrates that there is a performance different when $L_{\mathrm{cp}}$ changes, however, for sufficiently large $MN$ the capacities $C_{R_{\mathrm{cp}}}$ and $C_{R_{\mathrm{zp}}}$ can be approximated to
\begin{equation}
    C_{R_{\mathrm{cp}}} = C_{R_{\mathrm{zp}}} \approx \log _{2}\left(1+\gamma\right); ~~~~MN \gg\ 1.
\end{equation}
Whereas the capacities of FCP and FZS improve as the CP length value decreases.
Fig. \ref{fig:PE} shows the transmitted power needed to achieve a specific capacity assuming that $E\left\{|s(t)|^{2} \right\} = 1$. For $N=32$ the transmitted power for RCP, RZP, and FZS is shown to be not effect by $L_{\mathrm{cp}}$, unlike FCP scheme that transmits higher power than all other prefix/suffix schemes. This power is proportional to $L_{\mathrm{cp}}$.
As the subcarrier number $M$ increases, the transmitted power of all configurations converges to $P =\left(MN\right) E\left\{|s(t)|^{2} \right\}$.

\subsection{Fractional Doppler}\label{Subsec:frac}

\begin{table*}[t]
\caption{Comparison between the different prefix/suffix OTFS configuration.}
\centering
\begin{tabular}{l|c|c|c|c|}
\cline{2-5}
\multicolumn{1}{c|}{}                              & \textbf{\ac{RCP}} & \textbf{\ac{RZP}} & \textbf{\ac{FCP}} & \textbf{\ac{FZS}} \\ \hline
\multicolumn{1}{|l|}{\textbf{Phase Term $\Gamma^{(i)}$}}    & $e^{ j 2 \pi\frac{ k_i}{N}\frac{ l-l_i}{M}} \begin{cases}1 & l_{i} \leq l<M \\  e^{-\frac{j 2 \pi k}{N}} & 0 \leq l<l_{i}\end{cases}$            &      $e^{ j 2 \pi\frac{ k_i}{N}\frac{ l-l_i}{M}} \begin{cases}1 & l_{i} \leq l<M \\  e^{-\frac{j 2 \pi k}{N}} & 0 \leq l<l_{i}\end{cases}$               &   $e^{\frac{j2\pi k_i (L_{\mathrm{cp}}+l-l_i)}{(M+L_{\mathrm{cp}})N}}$               &   $e^{ j 2 \pi\frac{ k_i}{N}\frac{ l-l_i}{M}}$               \\ \hline
\multicolumn{1}{|l|}{$\Gamma^{(i)}$ in \textbf{Static Channel}}      & $ \begin{cases}1 & l_{i} \leq l<M \\  e^{-j 2 \pi\frac{k}{N}} & 0 \leq l<l_{i}\end{cases}$            &      $ \begin{cases}1 & l_{i} \leq l<M \\  e^{-j 2 \pi\frac{k}{N}} & 0 \leq l<l_{i}\end{cases}$               &   $1$               &   $1$                   \\ \hline
% \multicolumn{1}{|l|}{\textbf{Channel Estimation Complexity}}  &         Same            &     Same                &         Same         &         Same         \\ \hline
% \multicolumn{1}{|l|}{\textbf{Symbol Detection Complexity}}    &         Same            &      Same               &         Same         &         Same        \\ \hline
\multicolumn{1}{|l|}{\textbf{Power Efficiency} $E\left\{|s(t)|^{2} \right\} = 1$}    &         $\frac{1}{MN+L_{\mathrm{cp}}}$            &      $\frac{1}{MN}$               &        $\frac{1}{N(M+L_{\mathrm{cp}})}$           &      $\frac{1}{MN}$              \\ \hline
\multicolumn{1}{|l|}{\textbf{Spectral Efficiency}} &      $\frac{MN}{MN+L_{\mathrm{cp}}}$               &        $\frac{MN}{MN+L_{\mathrm{cp}}}$             &        $\frac{M}{M+L_{\mathrm{cp}}}$          &       $\frac{M-L_{\mathrm{cp}}}{M}$          \\ \hline
\end{tabular}
\label{tab:summary}
\end{table*}

One advantage of using full prefix rather than reduced prefix/suffix can be extracted from \eqref{equ:y_F}, where we see that the Doppler resolution is a function of the CP length $L_{\mathrm{cp}}$. Thus, not only FCP has higher Doppler resolution compared to \ac{RCP}/ZP OTFS but also the Doppler resolution itself is adaptive and can be controlled by changing the $L_{\mathrm{cp}}$ to achieve less inter-Doppler interference in the system. 
\begin{figure} [h]
\centering
\begin{subfigure}{.24\textwidth}
  \centering
  % include second image
  \includegraphics[width=1.05\linewidth]{ 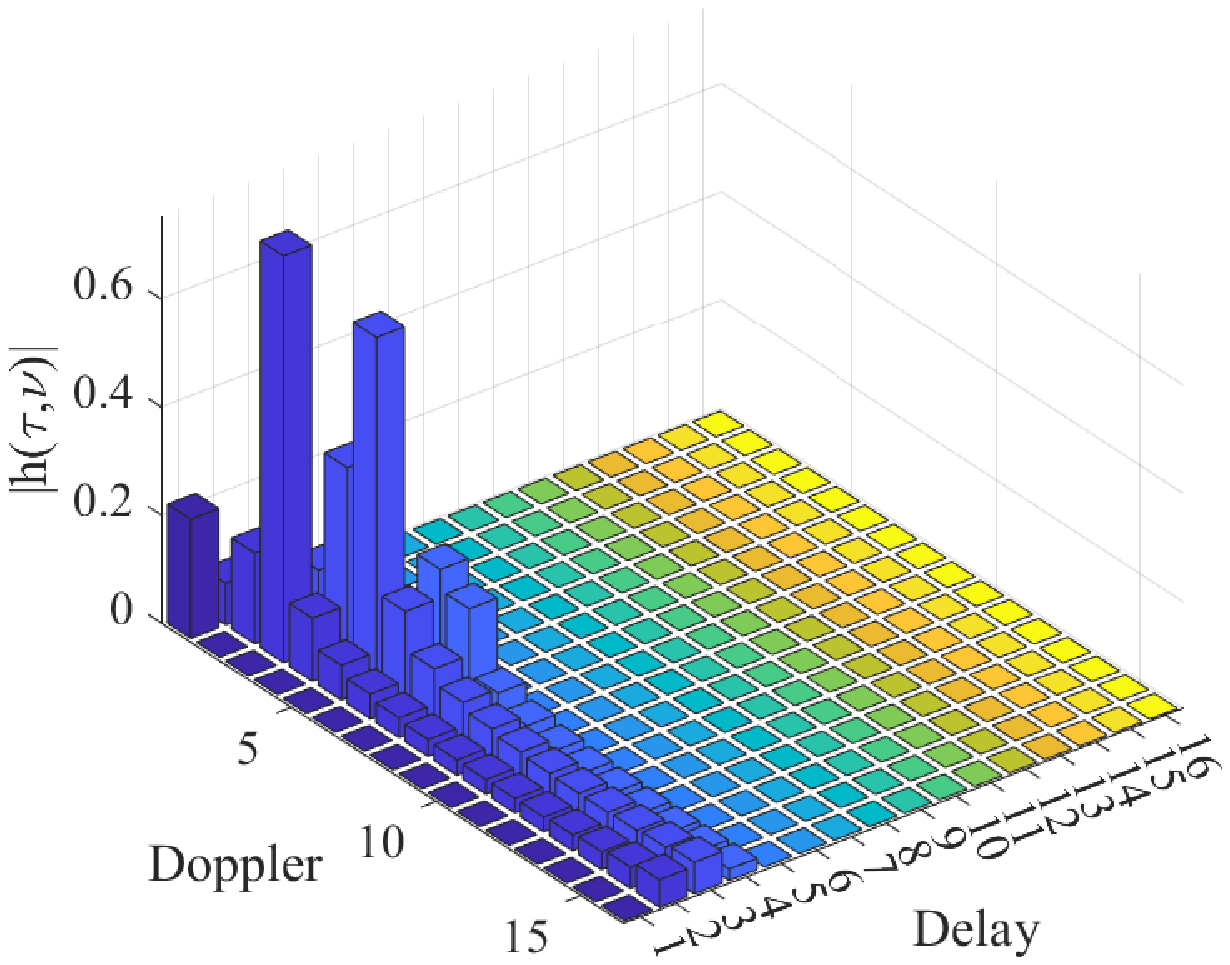}  
  \caption{$ L_{\mathrm{cp}} = 4$.}
%   \label{fig:sub-second}
\end{subfigure}
\begin{subfigure}{.24\textwidth}
  \centering
  % include third image
  \includegraphics[width=1.05\linewidth]{ 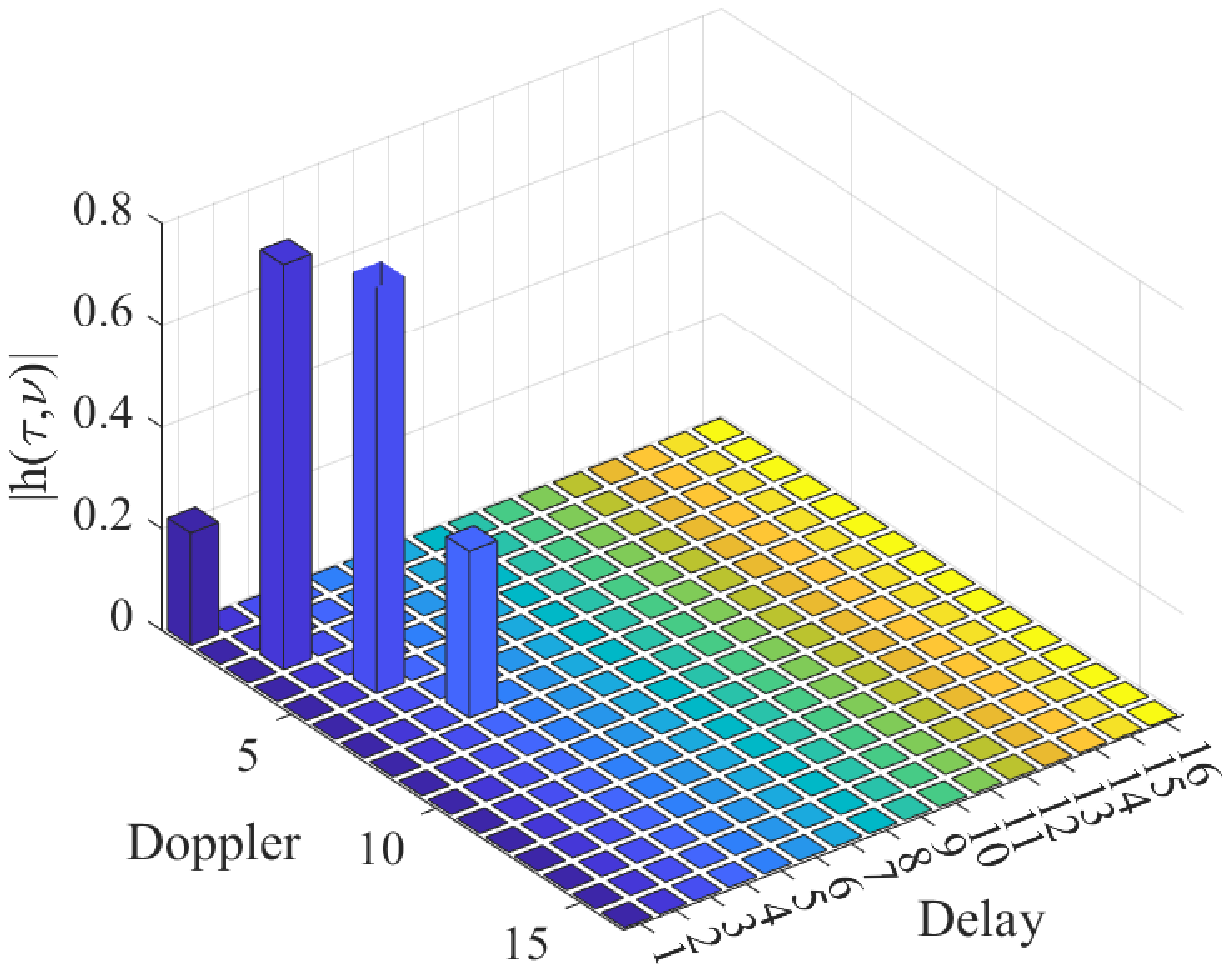}  
  \caption{$ L_{\mathrm{cp}} = 6$.}
%   \label{fig:sub-third}
\end{subfigure}
 \caption{The delay-Doppler \ac{CIR} of \ac{FCP} OTFS under fractional Doppler using different CP lengths. The number of paths in the delay-Doppler domain is $L = 4$.}
    \label{fig:CP_dep}
\end{figure}

Fig. \ref{fig:CP_dep} depicts the delay-Doppler \ac{CIR} in the existence of fractional Doppler. it is observed that as $L_{\mathrm{cp}}$ changes, the channel profile changes as well. For instance, when $L_{\mathrm{cp}}=4$ the channel suffers from high inter-Doppler interference which vanishes at $L_{\mathrm{cp}}=6$. 
Fig. \ref{fig:fra} shows that fractional Doppler kills the \ac{BER} performance of RCP, RZP, and FZS. However, in case of FCP, the BER performance changes as the CP length changes and achieves the best performance at $L_{\mathrm{cp}}=6$ emphasizing the results depicted in Fig. \ref{fig:CP_dep}. Table \ref{tab:summary} summarizes the main differences between different OTFS implementations.

\begin{figure}[h!]
	\centering
	\includegraphics[scale=0.50]{ 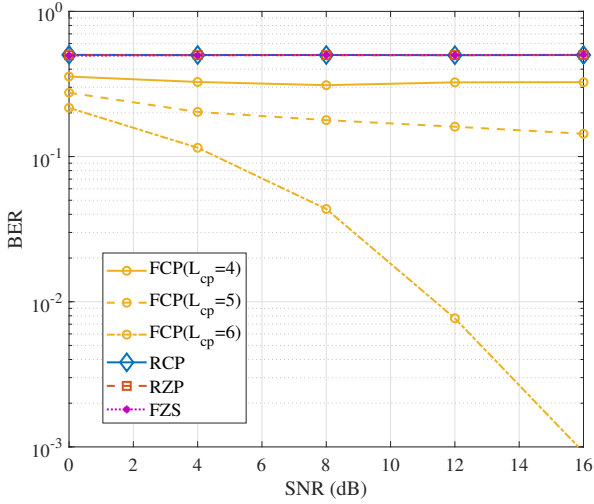}
    	\footnotesize\caption{The BER performance vs SNR of OTFS under fractional Doppler using different OTFS configurations and different CP lengths. The number of paths in the delay-Doppler domain is $L = 4$.}
	\label{fig:fra}
\end{figure}
\subsection{Reduced or Full?}\label{Subsec:structure}

The full prefix/suffix is different from the reduced one in terms of interpretation even though they mainly share the same task of combating channel effects. \ac{RCP} OTFS can resemble to \ac{CP}-OFDM, where the CP is appended to the beginning of the frame; however, most importantly this CP has no meaning on its own since it is generated from the combination of the whole data symbols, which is the reason why it is discarded the receiver side \cite{zegrar2022common}.
On the other hand, the CP in \ac{FCP} OTFS is interpretable and conveys data which can be extracted and decoded. Consider $\mathbf{P}^{\mathrm{F_{\mathrm{cp}}}} \in \mathbb{C}^{(M+L_{\mathrm{cp}})\times M}$ the CP addition matrix \cite{farhang2017low}, the 2D transmit signal can be given as follows
\begin{equation}
\begin{aligned}
    \mathbf{S}^{\mathrm{F_{\mathrm{cp}}}} & = \mathbf{P}^{\mathrm{F_{\mathrm{cp}}}} \mathbf{S} = \mathbf{P}^{\mathrm{F_{\mathrm{cp}}}} \mathbf{X}_{\mathrm{DD}}\mathbf{F}_N^H \\
    &= \mathbf{X}_{\mathrm{DD}}^{\mathrm{F_{\mathrm{cp}}}}\mathbf{F}_N^H,
\end{aligned}
\end{equation}
where $\mathbf{X}_{\mathrm{DD}}^{\mathrm{F_{\mathrm{cp}}}} = \mathbf{P}^{\mathrm{F_{\mathrm{cp}}}}\mathbf{X}_{\mathrm{DD}}$ represents the extended OTFS frame in the delay-Doppler domain. Therefore, adding \ac{FCP} is equivalent to extending the delay-Doppler grid to $(M+L_{\mathrm{cp}})\times N$ by repeating the last $L_{\mathrm{cp}}$ rows of $\mathbf{X}_{\mathrm{DD}}$. Also, this explains why the Doppler resolution in \eqref{equ:Dopp-res} is $(M+L_{\mathrm{cp}})\times N$ instead of $M\times N$ as in \ac{RCP} OTFS. Fig. \ref{fig:FCP} graphically illustrates the relationship between adding the CP in delay-Doppler and time domains in \ac{FCP} OTFS system.\footnote{Note that the same conclusion is also valid for \ac{FZS} OTFS.} 
So, spending more valuable resources to transmit repeated data, which could be replaced with unknown data symbols seems futile, knowing that one CP is enough to protect the whole OTFS frame and that the computational complexity and the performance of channel estimation/equalization and data detection are the same.

Therefore, we conclude that the use of \ac{FCP} OTFS should be well motivated otherwise \ac{RCP} OTFS is more promising to be used. For instance, one advantage of \ac{FCP} over \ac{RCP} is mentioned in Subsection \ref{Subsec:Fcp} where if the channel is static, the input-output relation of \ac{FCP} OTFS becomes 2D circular convolution between the data and the delay-Doppler \ac{CIR} as given in \eqref{equ:yy}, which is not the case for \ac{RCP} OTFS. In this case, using \textit{Theorem 1} enables the linear channel equalization of OTFS as in OFDM systems using \eqref{equ:lin}. Another advantage of \ac{FCP} OTFS is the block-diagonal structure of the channel matrix in time which allows simple channel equalization in \ac{MIMO}-OTFS systems \cite{qu2022efficient}.

\subsection{Reduced-\ac{FCP} OTFS}

In this subsection, a novel OTFS CP configuration is proposed in which one \ac{RCP} is appended to the \ac{FCP} OTFS frame as depicted in Fig. \ref{fig:rfcp_conf}. We call this type of mixed CP configuration RFCP OTFS. Doing so, we ensure that both the symbols in the data block ($M\times N$ symbols) and symbols in the CP block ($N\times L_{\mathrm{cp}}$ symbols) can be decoded. Specifically at the receiver side, the \ac{RCP} is discarded, then the received samples are reshaped as follows
\begin{equation}
Y(l,k)= \begin{cases} Y^{R_{\mathrm{cp}}}(l,k), & \text { if } l\leq M   \\ 
Y^{F_{\mathrm{cp}}}(l,k), & \text { if } l> M \end{cases}.
\label{equ:RFCP}
\end{equation}
From \eqref{equ:RFCP} it is seen that the received delay-Doppler data signal will follow the model in \eqref{equ:R_cont} where the repeated data in the \ac{FCP} block will follow the relation in \eqref{equ:y_F}. 

\begin{figure}[!h]
	\centering
 	\setcounter{figure}{7}
	\includegraphics[scale=0.80]{ 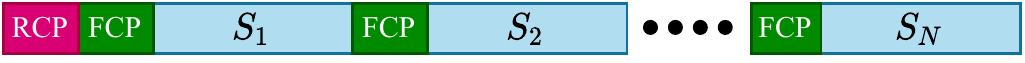}
    	\footnotesize\caption{Reduced-full CP OTFS frame.}
	\label{fig:rfcp_conf}
\end{figure}

\begin{figure*}[h]
	\centering
 	\setcounter{figure}{6}
	\includegraphics[scale=0.85]{ 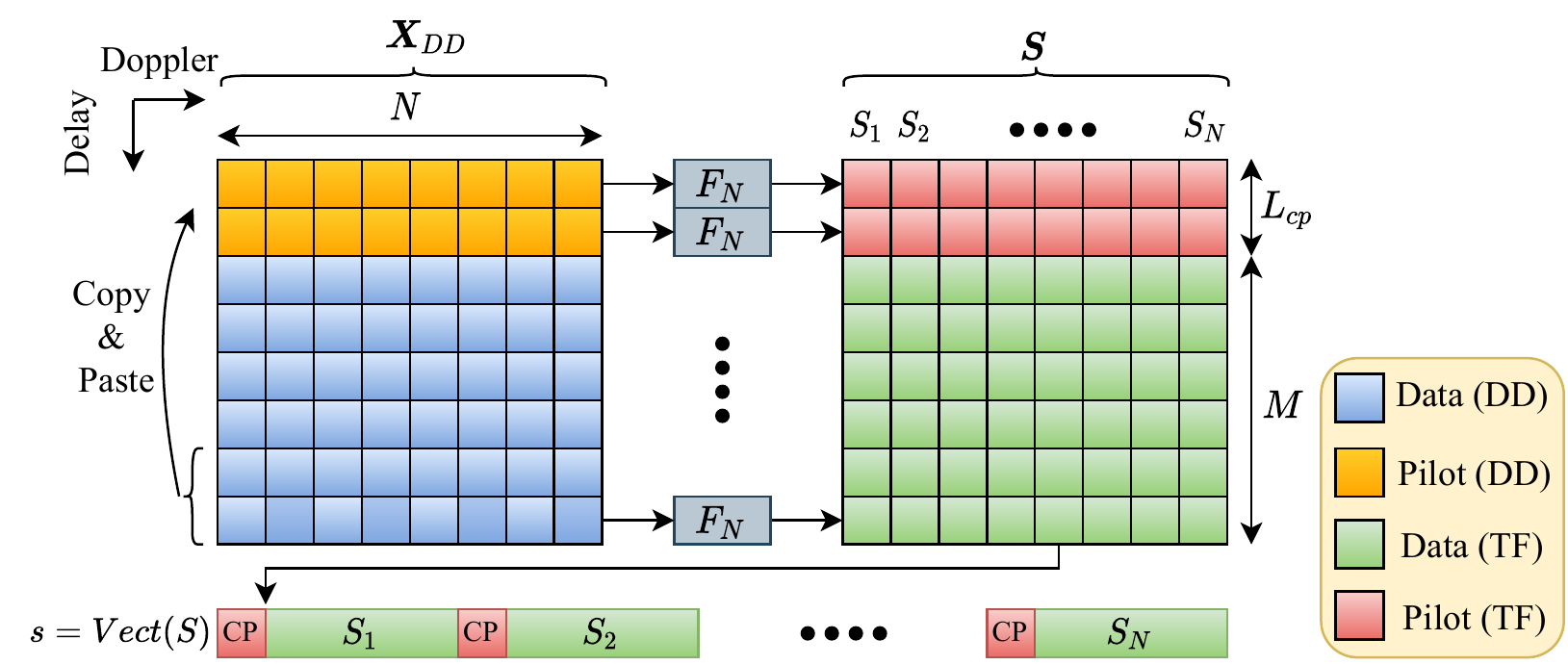}
    	\footnotesize\caption{The CP addition in \ac{FCP} OTFS in both delay-Doppler and time domains.}
	\label{fig:FCP}
\end{figure*}

The RFCP OTFS configuration allows the exploitation of the CPs in OTFS unlike OFDM as explained in Subsection \ref{Subsec:structure}. For instance, the repeated data in the \ac{FCP} can be used to provide diversity between the transmitted data symbols. Also, it can be used to reduce the guard interval dedicated for pilot \cite{raviteja2019embedded} into half by locating the pilot at the end part of the delay-Doppler grid. An illustration is shown in Fig. \ref{fig:double-CP} where the CP block in FCP OTFS is compared to the one in the proposed RFCP frame. For that, one pilot is inserted at the end of the delay index (i.e., $l = M-1$) so that it is copied in the full CP portion.
After passing through the doubly dispersive channel, it is seen that the pilot spreads accordingly with the propagation environment which has $L=4$ taps.
At the receiver side, the data block and the CP block in both FCP and RFCP OTFS systems are plotted and shown in Fig. \ref{fig:double-CP}. It is observed that the pilot in the data block spreads similarly in both configurations following the model in \eqref{equ:R_cont} as depicted in Fig. \ref{fig:double-CP} (a). On the other hand, it is seen that the CP block in the FCP frame suffers from high inter-Doppler interference as shown in Fig. \ref{fig:double-CP} (b). On the contrary, Fig. \ref{fig:double-CP} (c) shows an intact \ac{CIR} following the model in \eqref{equ:R_cont} where the appended reduced CP provides enough protection to secure the CP block in RFCP OTFS.

\begin{figure} []
\setcounter{figure}{8}
\centering
\begin{subfigure}[b]{0.402\textwidth}
   \includegraphics[width=0.9\linewidth]{ 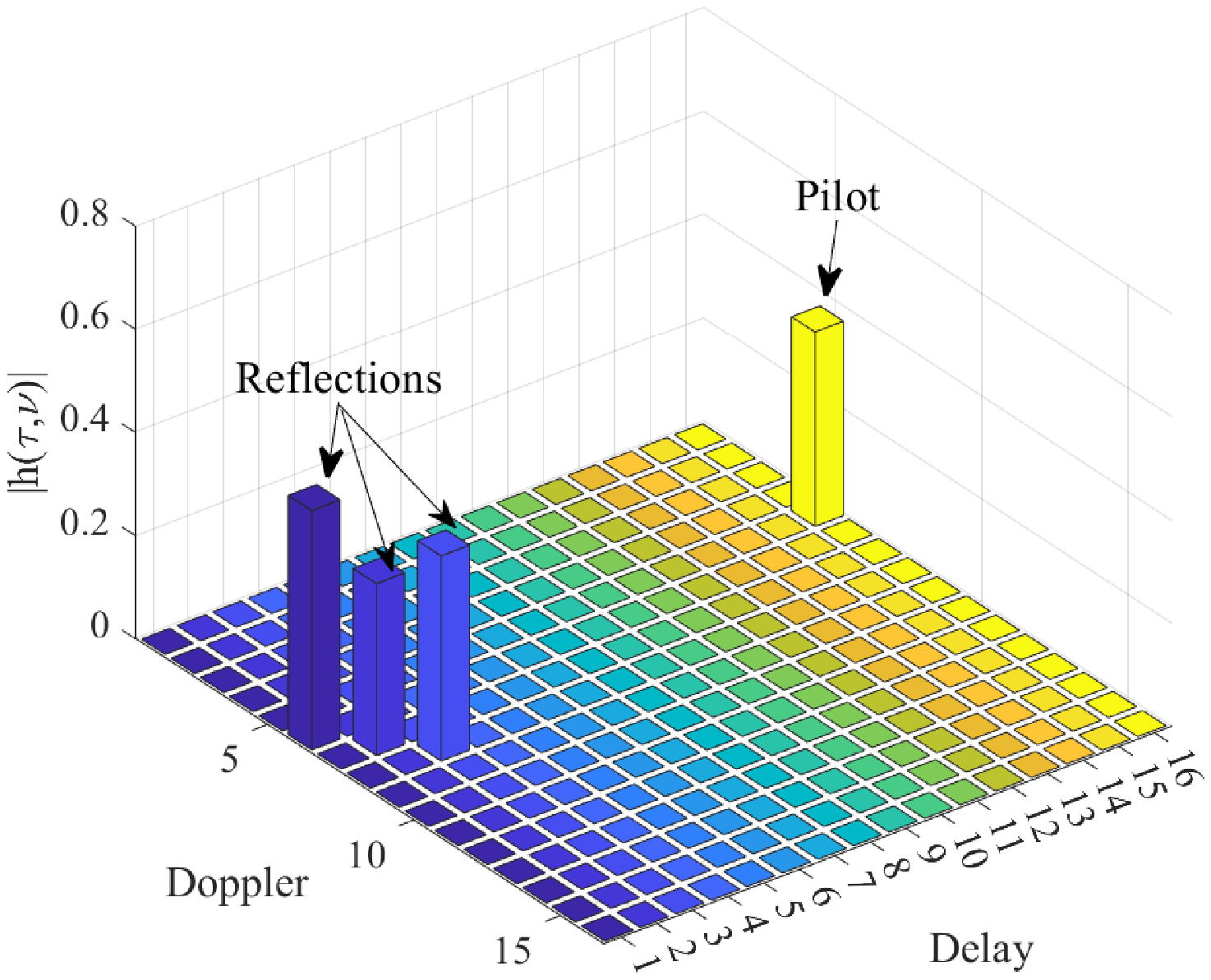}
   \caption{The received signal after discarding CPs.}
\end{subfigure}
\newline
\begin{subfigure}{.24\textwidth}
  \centering
  \includegraphics[width=0.9\linewidth]{ 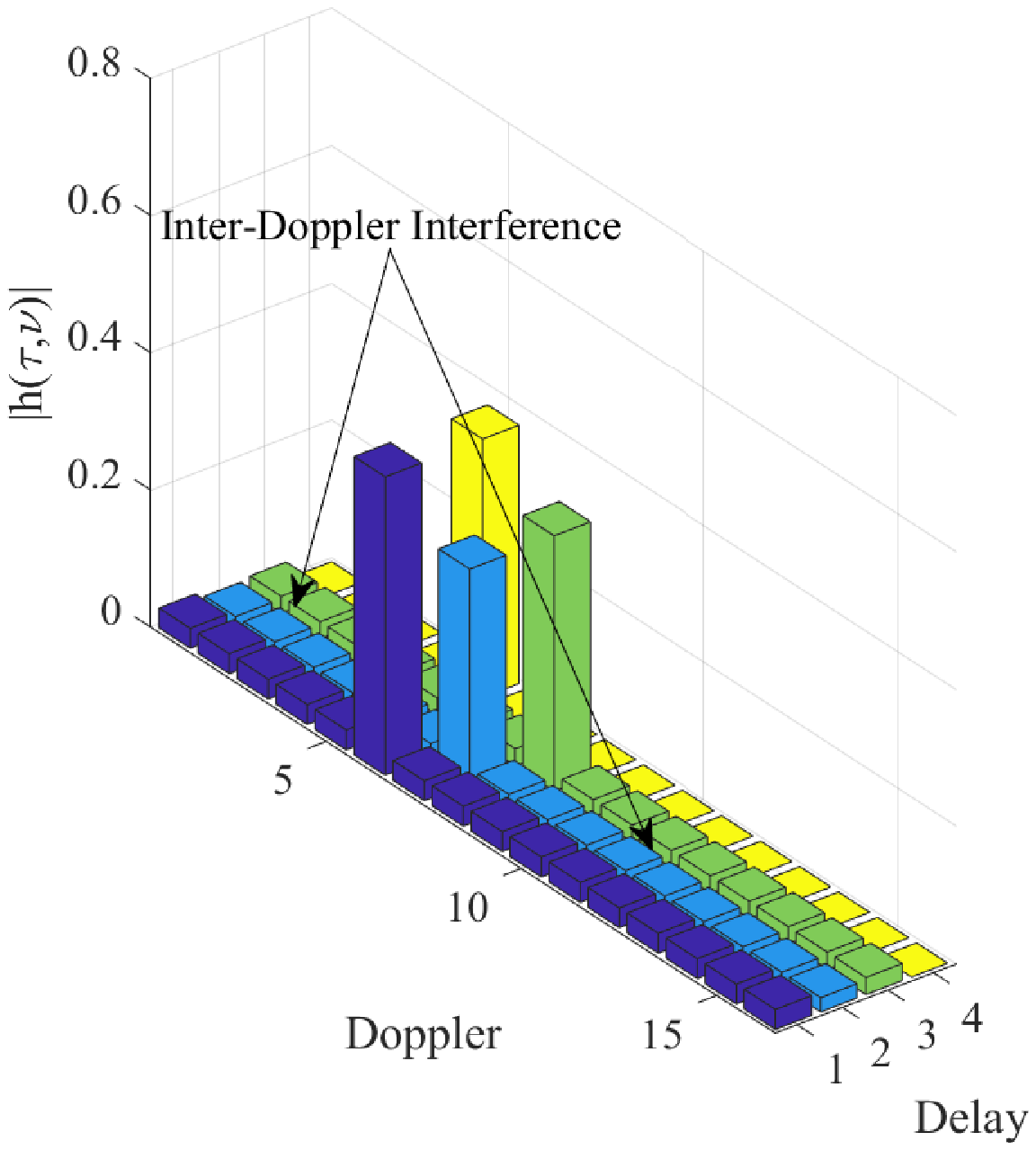}  
  \caption{The discarded CPs in FCP-OTFS.}
  \label{fig:sub-second}
\end{subfigure}
\begin{subfigure}{.24\textwidth}
  \centering
  \includegraphics[width=0.9\linewidth]{ 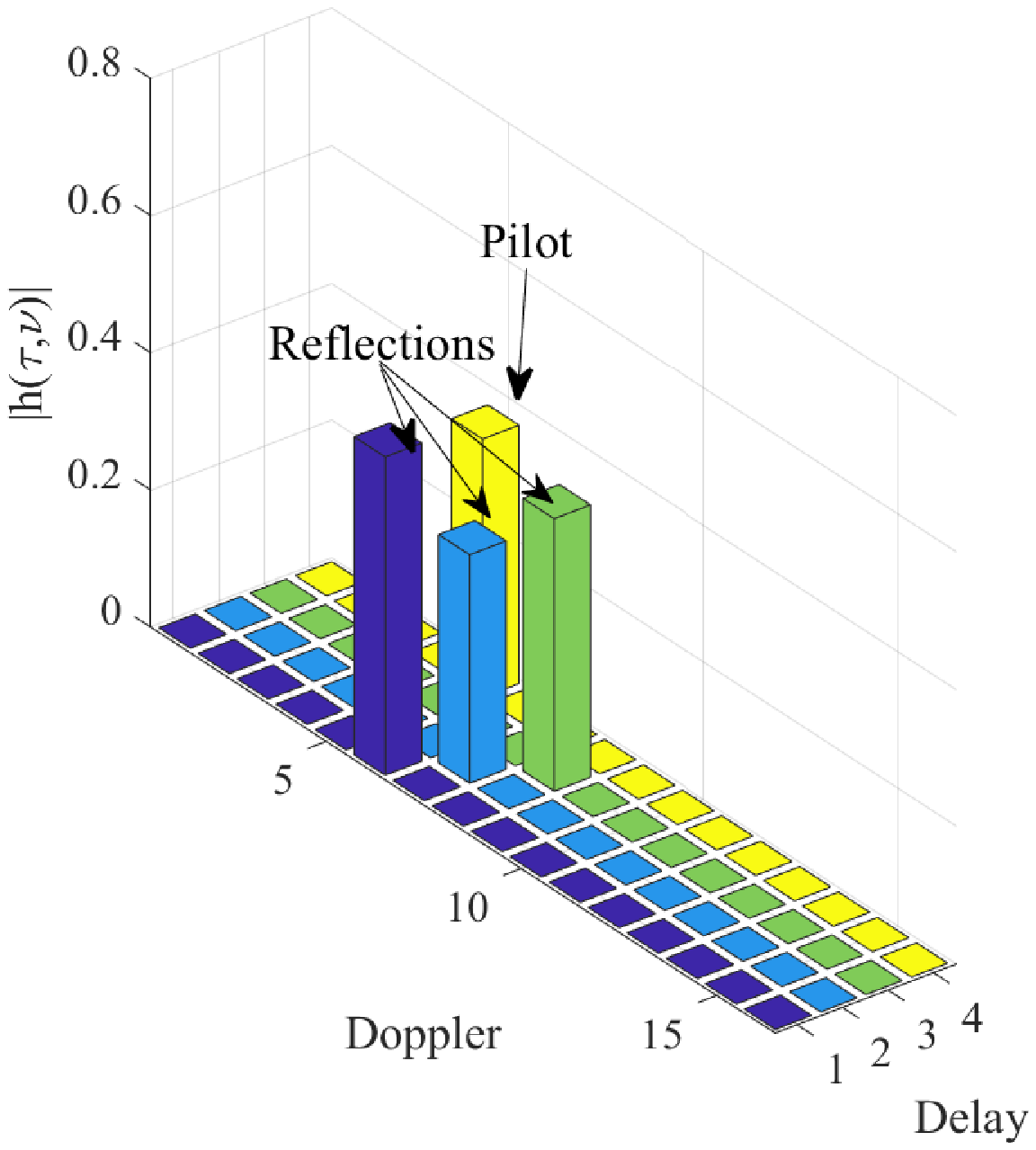}  
  \caption{The discarded CPs in RFCP-OTFS.}
  \label{fig:sub-third}
\end{subfigure}
\caption{ The delay-Doppler \ac{CIR} of RFCP OTFS using $(M;N;L_{\mathrm{cp}})=(16,16,4)$. The number of paths in the delay-Doppler domain is $L = 4$.}\label{fig:double-CP}
\end{figure}
%%%%%%%%%%%%%%%%%%%%%%%%%%%%%%%%%%%%%%%%%%
% ========================================================
\section{Conclusion} \label{section:Conclusion}

In this paper, the input-output relation of OTFS systems has been analyzed when considering rectangular pulse shaping for \ac{RCP}, \ac{RZP}, \ac{FCP}, and \ac{FZS} prefixes and suffixes. It is shown that regardless of the prefix/suffix type used, the received signal model is similar except for the phase value corresponding the quasi-periodic shift in delay-Doppler domain. Consequently, it is concluded that channel estimation complexity, symbol detection performance are the same for all OTFS configurations. A novel mixed prefix type has been proposed where jointly reduced and full CP structures are appended to the OTFS frame, namely RFCP OTFS. In RFCP OTFS systems, the CP blocks are not discarded, and are being decoded along with data frame at the receiver side. Based on the application/need, the most appropriate prefix/suffix configuration has been defined to be adopted in the system model. In our future work, we will consider the effect of both fractional delay and Doppler in our model and accordingly we will analyze the different OTFS realizations under the new constraints.

% ========================================================
\begin{appendices}
% ========================================================
\section{Proof of \textit{Theorem 1}}\label{App:Double_Circulant}

A doubly-block circulant matrix is the outcome of the Kronecker product of two circulant matrices \cite{calbert2020learning}. Let $\mathbf{C}_{1} \in \mathbb{C}^{N \times N}, \mathbf{C}_{2} \in \mathbb{C}^{M \times M}$ be two circulant matrices and $\mathbf{A} \in \mathbb{C}^{M N \times M N}$, is the matrix resulting from the Kronecker product of $\mathbf{C}_{1}$ and $\mathbf{C}_{2}$ as follows
\begin{equation}
    \begin{aligned}
    \mathbf{A} & =\mathbf{C}_{1} \otimes \mathbf{C}_{2} . 
    \end{aligned}
\end{equation}
Since $\mathbf{C}_{1}$ and $\mathbf{C}_{2}$ are circulant then they can be diagonalized using \ac{DFT} matrices i.e., $\mathbf{F}_N^{-1} \mathbf{D}_{1} \mathbf{F}_N$ and $\mathbf{F}_{M} \mathbf{D}_{2} \mathbf{F}_{M}^{-1}$ \cite{davis1994circulant}. Then,
\begin{equation}
    \begin{aligned}
    \mathbf{A} & =\left(\mathbf{F}_N^{-1} \mathbf{D}_{1} \mathbf{F}_N\right) \otimes\left(\mathbf{F}_{M} \mathbf{D}_{2} \mathbf{F}_{M}^{-1}\right) \\
    & =\left(\mathbf{F}_{N}^{-1} \otimes \mathbf{F}_{M}\right)\left(\mathbf{D}_{1} \otimes \mathbf{D}_{2}\right)\left(\mathbf{F}_{N} \otimes \mathbf{F}_{M}^{-1}\right) . 
    \end{aligned}
\end{equation}
Note that $\mathbf{D}_{1} \otimes \mathbf{D}_{2}$ is a diagonal matrix because the Kronecker product of two diagonal matrices is a diagonal matrix. Therefore,
\begin{equation}
    \boldsymbol{\Sigma}=\left(\mathbf{D}_{1} \otimes \mathbf{D}_{2}\right) =\left(\mathbf{F}_{N} \otimes \mathbf{F}_{M}^{-1}\right)\mathbf{A}\left(\mathbf{F}_{N}^{-1} \otimes \mathbf{F}_{M}\right) . 
\end{equation}
%================================
\section{Proof of \eqref{equ:R_cont}}\label{App:y_Rcp}
Each received element in the \ac{RCP} system is given as 
 \begin{equation}
 \begin{aligned}
 Y^{\mathrm{R_{\mathrm{cp}}}}(l, k)&=\sum_{\beta=0}^{MN-1}\mathbf{H}_{\mathrm{eff}}^{\mathrm{R_{\mathrm{cp}}}}(kM+l,\beta) \mathbf{x}(\beta)\\
&=\sum_{l^{\prime}=0}^{M-1} \sum_{k^{\prime}=0}^{N-1} \mathbf{H}_{\mathrm{eff}}^{\mathrm{R_{\mathrm{cp}}}}(kM+l, k^\prime M+l^\prime) X_{\mathrm{DD}}(l^\prime, k^\prime)\\
&=\sum_{l^{\prime}=0}^{M-1} \sum_{k^{\prime}=0}^{N-1}X_{\mathrm{DD}}(l^\prime, k^\prime)\sum_{i=0}^{L-1}h_i\mathbf{T}^{(i)}(kM+l, k^\prime M+l^\prime).
 \end{aligned}
 \label{equ:temp1}
\end{equation}
Note that $\mathbf{T}^{(i)}(kM+l, k^\prime M+l^\prime)$ is non-zeros only when $l^\prime-[l-l_{i}]_M=k^\prime-[k-k_{i}]_N=0$. Then, by substituting \eqref{equ:T} in \eqref{equ:temp1}, we find
 \begin{equation}
 \begin{aligned}
 &Y^{\mathrm{R_{\mathrm{cp}}}}(l, k)=\sum_{l^{\prime}=0}^{M-1} \sum_{k^{\prime}=0}^{N-1}X_{\mathrm{DD}}(l^\prime, k^\prime)\sum_{i=0}^{L-1}h_i\Lambda_i(l, k)e^{ j 2 \pi\frac{ k_i}{N}\frac{ l-l_i}{M}}\\
  &~~~~~~~~~~~~~~~~~\cdot\delta \left( l^\prime-[l-l_{i}]_M\right)
 \delta \left( k^\prime-[k-k_{i}]_N\right)\\
 & = \sum_{i=0}^{L-1} h_{i}e^{ j 2 \pi\frac{ k_i}{N}\frac{ l-l_i}{M}} \Lambda_i(l,k) 
 X_{\mathrm{DD}}\left(\left[l-l_{i}\right]_{M},\left[k-k_{i}\right]_{N}\right),
 \end{aligned}
\end{equation}

%where $w(l, k)$ denotes \ac{AWGN}. 
where 
\begin{equation}
\Lambda_i(l, k)= \begin{cases}1 & l_{i} \leq l<M \\  e^{-j 2 \pi\frac{k}{N}} & 0 \leq l<l_{i}\end{cases}.
\end{equation}
%================================
\section{Proof of \eqref{equ:G(i,j)}}\label{app:H(i,j)}
The permutation matrix $\boldsymbol{\Pi}^{l_{i}}_{M}$ is an identity matrix with columns shifted by a value of $l_{i}$. Then, each element of this matrix can be found as
\begin{equation}
    \boldsymbol{\Pi}^{l_{i}}_{M}(l^\prime,k^\prime) = \delta \left( [l^\prime-k^\prime]_M-l_{i}\right).
\end{equation}
Since $\boldsymbol{\Delta}^{k_{i},n}_{M}$ is a diagonal matrix, the $(l^\prime,k^\prime)$-th element of the of $\mathbf{H}_n^{\mathrm{F_{\mathrm{cp}}}}(l^\prime,k^\prime)$ is calculated as
\begin{equation}
\begin{aligned}
    &\mathbf{H}_n^{\mathrm{F_{\mathrm{cp}}}}(l^\prime,k^\prime) = \sum_{i=0}^{L-1} h_{i}\boldsymbol{\Delta}^{k_{i},n}_{M}\left( [l^\prime-l_{i}]_M,k^\prime\right)\\
    &=\sum_{i=0}^{L-1}h_{i}z_i^{\left(n(M+L_{\mathrm{cp}})+L_{\mathrm{cp}}+l^\prime-l_i\right)} \cdot \delta \left( [l^\prime-k^\prime]_M-l_{i}\right).
\end{aligned}
    \label{equ:pi}
\end{equation}
substituting \eqref{equ:pi} in \eqref{equ:G(i,j)} we find

\begin{equation}
\begin{aligned}
    \mathbf{G}^{\mathrm{F_{\mathrm{cp}}}}_n(l^\prime,k^\prime) &= \sum_{\alpha=0}^{N}\sum_{i=0}^{L-1}h_{i}z_i^{\left(\alpha(M+L_{\mathrm{cp}})+L_{\mathrm{cp}}+l^\prime-l_i\right)}e^{-j2\pi nk/N}\\
    & \cdot \delta \left( [l^\prime-k^\prime]_M-l_{i}\right).
\end{aligned}
\end{equation}

%================================
\section{Proof of \eqref{equ:y_F}}\label{App:y_F}
The vectorized form of the received signal in \eqref{equ:y_F} can be written as
\begin{equation}
 \mathbf{Y}^{\mathrm{F_{\mathrm{cp}}}} = \sum_{\alpha=0}^{N-1} \mathbf{H}_{\alpha} \mathbf{X}_{\mathrm{DD}} \mathbf{f}_{\alpha}^{*} \mathbf{f}_{\alpha}^{T}=\sum_{\alpha=0}^{N-1} \mathbf{H}_{\alpha} \mathbf{X}_{\mathrm{DD}} \boldsymbol{\Pi}_{\alpha},
\end{equation}
where $\boldsymbol{\Pi}_{\alpha}(l^\prime,k^\prime) = e^{j2\pi \alpha\frac{l^\prime-k^\prime}{N}}$.
\begin{equation}
\begin{aligned}
Y^{\mathrm{F_{\mathrm{cp}}}}(l, k)&=\sum_{\alpha=0}^{N-1} \sum_{l^{\prime}=0}^{M-1} \sum_{k^{\prime}=0}^{N-1} H_{\alpha}^{\mathrm{F_{\mathrm{cp}}}}\left[l, l^{\prime}\right] X_{\mathrm{DD}}\left[l^{\prime}, k^{\prime}\right] \Omega_{\alpha}\left[k^{\prime},k\right]\\
&=\sum_{\alpha=0}^{N-1} \sum_{l^{\prime}=0}^{M-1} \sum_{k^{\prime}=0}^{N-1}\bigg[\sum_{i=0}^{L-1}h_{i}z_i^{\left(\alpha(M+L_{\mathrm{cp}})+L_{\mathrm{cp}}+l-l_i\right)} \bigg.\\
&~~~~~~~~~~~~~~\bigg.\cdot \delta \left( [l-l^\prime]_M-l_{i}\right)\bigg] X_{\mathrm{DD}}\left[l^{\prime}, k^{\prime}\right] e^{j2\pi \alpha\frac{k^\prime-k}{N}}\\
&=\sum_{l^{\prime}=0}^{M-1} \sum_{k^{\prime}=0}^{N-1} X_{\mathrm{DD}}\left[l^{\prime}, k^{\prime}\right] \sum_{i=0}^{L-1}h_{i} \cdot \delta \left( [l-l^\prime]_M-l_{i}\right)\\
&~~~~~~~~~~~\times \sum_{\alpha=0}^{N-1} e^{j2\pi k_i\frac{\left(\alpha(M+L_{\mathrm{cp}})+L_{\mathrm{cp}}+l-l_i\right)}{(M+L_{\mathrm{cp}})N}} e^{j2\pi \alpha\frac{k^\prime-k}{N}}\\
&=\sum_{l^{\prime}=0}^{M-1} \sum_{k^{\prime}=0}^{N-1} X_{\mathrm{DD}}\left[l^{\prime}, k^{\prime}\right] \sum_{i=0}^{L-1}h_{i} \cdot \delta \left( [l-l^\prime]_M-l_{i}\right) \\
&~~~~~~~~~~~~~~\times e^{\frac{2\pi k_i (L_{\mathrm{cp}}+l-l_i)}{(M+L_{\mathrm{cp}})N}} \sum_{\alpha=0}^{N-1}  e^{j2\pi \alpha\frac{k_i+(k^\prime-k)}{N}}.\\
\end{aligned}
\end{equation}
In the absence of fractional Doppler $\sum_{\alpha=0}^{N-1}  e^{j2\pi \alpha\frac{k_i+(k^\prime-k)}{N}} = \delta \left( [k-k^\prime]_N-k_i\right)$. Then,
\begin{equation}
\begin{aligned}
Y^{\mathrm{F_{\mathrm{cp}}}}(l, k)&=\sum_{l^{\prime}=0}^{M-1} \sum_{k^{\prime}=0}^{N-1} X_{\mathrm{DD}}\left[l^{\prime}, k^{\prime}\right] \sum_{i=0}^{L-1}h_{i} \cdot \delta \left( [l-l^\prime]_M-l_{i}\right)\\
&~~~~~~~~~~~~~~\delta \left( [k-k^\prime]_N-k_i\right) e^{\frac{j2\pi k_i (L_{\mathrm{cp}}+l-l_i)}{(M+L_{\mathrm{cp}})N}}\\
&=\sum_{i=0}^{L-1} h_{i}e^{\frac{j2\pi k_i (L_{\mathrm{cp}}+l-l_i)}{(M+L_{\mathrm{cp}})N}} X_{\mathrm{DD}}\left(\left[k-k_{i}\right]_{N},\left[l-l_{i}\right]_{M}\right).
\end{aligned}
\end{equation}

%================================
\section{Proof of \eqref{equ:G(i,jj)}}\label{App:G_zs}
The elements of the lower triangular matrix of $\mathbf{L}_\alpha^{\mathrm{F_{\mathrm{zs}}}}$ can be computed as follows
\begin{equation}
\begin{aligned}
    &\mathbf{L}_\alpha^{\mathrm{F_{\mathrm{zs}}}}(l^\prime,k^\prime) = \sum_{i=0}^{L-1} h_{i}\boldsymbol{\Delta}^{k_{i},n}_{M}\left( l^\prime-l_{i},k^\prime\right)\\
    &=\sum_{i=0}^{L-1}h_{i}z_i^{\left(\alpha M+l-l_{i}\right)} \cdot \delta \left( l^\prime-k^\prime-l_{i}\right).
\end{aligned}
    \label{equ:pii}
\end{equation}
substituting \eqref{equ:pii} in \eqref{equ:G(ii,j)} we find

\begin{equation}
\begin{aligned}
    \boldsymbol{\Omega}^{\mathrm{F_{zs}}}_n(l^\prime,k^\prime) &= \sum_{\alpha=0}^{N}\sum_{i=0}^{L-1}h_{i}z_i^{(\alpha M +l-l_{i})}e^{-j2\pi nk/N}
     \cdot \delta \left( l^\prime-k^\prime-l_{i}\right).
\end{aligned}
\end{equation}

%================================
\section{Proof of \eqref{equ:y_Fzs}}\label{App:y_Fzs}
To derive the expression of the received signal $Y^{\mathrm{F_{zs}}}(l, k)$ in \eqref{equ:y_Fzs}, we can follow the same steps in Appendix \ref{App:y_F} by considering $L_{\mathrm{cp}} = 0$ and $\mathbf{L}^{\mathrm{F_{zs}}}_\alpha$ instead of $\mathbf{H}^{\mathrm{F_{\mathrm{cp}}}}_\alpha$. Then, we find

\begin{equation}
\begin{aligned}
Y^{\mathrm{F_{zs}}}(l, k)&=\sum_{\alpha=0}^{N-1} \sum_{l^{\prime}=0}^{M-1} \sum_{k^{\prime}=0}^{N-1} L_{\alpha}^{\mathrm{F_{\mathrm{zs}}}}\left[l, l^{\prime}\right] X_{\mathrm{DD}}\left[l^{\prime}, k^{\prime}\right] \Omega_{\alpha}\left[k^{\prime},k\right]\\
&=\sum_{\alpha=0}^{N-1} \sum_{l^{\prime}=0}^{M-1} \sum_{k^{\prime}=0}^{N-1}\bigg[\sum_{i=0}^{L-1}h_{i}z_i^{\left(\alpha M +l^{\prime}\right)} \cdot \delta \left( l-l^\prime-l_{i}\right)\bigg] \\
&~~~~~~~~~~~~~~ \times X_{\mathrm{DD}}\left[l^{\prime}, k^{\prime}\right] e^{j2\pi \alpha\frac{k^\prime-k}{N}}\\
&=\sum_{l^{\prime}=0}^{M-1} \sum_{k^{\prime}=0}^{N-1} X_{\mathrm{DD}}\left[l^{\prime}, k^{\prime}\right] \sum_{i=0}^{L-1}h_{i} \cdot \delta \left( l-l^\prime-l_{i}\right)\\
&~~~~~~~~~~~~~~\delta \left( [k-k^\prime]_N-k_i\right) e^{\frac{j2\pi k_i l^\prime}{MN}}\\
&=\sum_{i=0}^{L-1} h_{i}e^{\frac{j2\pi k_i (l-l_{i})}{MN}} X_{\mathrm{DD}}\left(\left[k-k_{i}\right]_{N},l-l_{i}\right).
\end{aligned}
\end{equation}
Note that $l-l_{i}$ is not modulo $M$ anymore. This because $\mathbf{L}_{\alpha}^{\mathrm{F_{\mathrm{zs}}}}$ is lower triangular matrix as proved in Appendix \ref{App:G_zs}. 
\end{appendices}

% \bibliographystyle{IEEEtran}
% \bibliography{IEEEfull,Bibliography}

\vfill

\end{document}